\documentclass[amsmath,twocolumn,prb,nofootinbib,superscriptaddress,longbibliography]{revtex4-2} 
\usepackage{amsthm,amssymb,amsfonts,graphicx,verbatim, xcolor,bm} 
\usepackage{hyperref} 
\usepackage[utf8]{inputenc} 
\usepackage{siunitx} 

\usepackage{dsfont}

\usepackage{csquotes}
\MakeOuterQuote{"}

\allowdisplaybreaks

\begin{document}

\newtheorem{theorem}{Theorem}[section]
\newtheorem{corollary}{Corollary}[theorem]
\newtheorem{example}{Example}[theorem]
\newtheorem{lemma}[theorem]{Lemma}
\newtheorem{proposition}{Proposition}
\newcommand{\beqn}{\begin{eqnarray}}
\newcommand{\eeqn}{\end{eqnarray}}
\newcommand{\ee}{\epsilon}
\newcommand{\diag}{{\text{diag}}}
\newcommand{\tw}{\tilde{w}}
\newcommand{\br}{\bm r}
\newcommand{\bk}{\bm k}
\newcommand{\bq}{\bm q}
\newcommand{\eqn}[1]{(\eqref{#1})}
\newcommand{\bQ}{\bm Q}
\newcommand{\bG}{\bm G}
\newcommand{\bb}{\bm b}
\newcommand{\ba}{\bm a}
\newcommand{\bR}{\bm R}
\newcommand{\bp}{\bm p}
\newcommand{\mr}{moir\'e~}
\newcommand{\Mr}{Moir\'e~}
\newcommand{\mom}{supermoir\'e~}
\newcommand{\Mom}{Supermoir\'e~}
\newcommand{\bKM}{{\bf K}_\text{M}}
\newcommand{\bGM}{{\bf \Gamma}_\text{M}}

\title{Nature of even and odd magic angles in helical twisted trilayer graphene}

\author{Daniele Guerci}
\affiliation{Center for Computational Quantum Physics, Flatiron Institute, New York, New York 10010, USA}

\author{Yuncheng Mao}
\affiliation{Universit\'e Paris Cit\'e, CNRS,  Laboratoire  Mat\'eriaux  et  Ph\'enom\`enes  Quantiques, 75013  Paris,  France}

\author{Christophe Mora}
\affiliation{Universit\'e Paris Cit\'e, CNRS,  Laboratoire  Mat\'eriaux  et  Ph\'enom\`enes  Quantiques, 75013  Paris,  France}

\date{\today}

\begin{abstract}

Helical twisted trilayer graphene exhibits zero-energy flat bands with large degeneracy in the chiral limit. The flat bands emerge at a discrete set of magic twist angles and feature properties intrinsically distinct from those realized in twisted bilayer graphene. Their degeneracy and the associated band Chern numbers depend on the parity of the magic angles. Two degenerate flat bands with Chern numbers $C_A=2$ and $C_B=-1$ arise at odd magic angles, whereas even magic angles display four flat bands, with Chern number $C_{A/B}=\pm1$, together with a Dirac cone crossing at zero energy. All bands are sublattice polarized. We demonstrate the structure behind these flat bands and obtain analytical expressions for the wavefunctions in all cases.  Each magic angle is identified with the vanishing of a zero-mode wavefunction at high-symmetry position and momentum. The whole analytical structure results from whether the vanishing is linear or quadratic for the, respectively, odd and even magic angle. The $C_{3z}$ and $C_{2y}T$ symmetries are shown to play a key role in establishing the flat bands. In contrast, the particle-hole symmetry is not essential, except from gapping out the crossing Dirac cone at even magic angles.

\end{abstract}

\maketitle

\section{Introduction}

Twisted trilayer graphene (TTG) has recently received significant attention due to its structural similarities to twisted bilayer graphene while exhibiting distinct features. Different configurations of TTG have been considered based on the relative twist directions of the top and bottom layers:  (1) mirror symmetrical TTG~\cite{park2021tunable,kim2022evidence,kim2023imaging,Caligaru_2021,Xie_2021,Lei_2021,Christos_PRX_2022,Daniele_2022_mTTG}, where the top and bottom layers are rotated in the same direction by the same angle relative to the middle layer, (2) twisted monolayer-bilayer graphene~\cite{Chen_2020,He_2021,Polshyn_2021,Ledwith_Khalaf_2022,yang2023flat,Morell_2013_tri}, where Bernal stacked bilayer graphene is rotated with a small angle relative to the third layer, and (3) helical TTG~\cite{uri2023superconductivity,Christophe_2019,PhysRevLett.125.116404,Yuncheng2023,guerci2023chern,devakul2023magicangle,nakatsuji2023multiscale,foo2023extended}, where the top and bottom layers are rotated in opposite directions. By adding an additional graphene layer on top of the bilayer, TTG offers additional "knobs" to manipulate the system's physical properties. Besides the twist angle, the layer shifting and displacement field have also been identified as key factors for altering the physical properties of TTG. These features open up promising avenues for studying and controlling the unique electronic properties of TTG systems.

Recent theoretical studies~\cite{Yuncheng2023,Dunbrack2013,popov2023magic,guerci2023chern,devakul2023magicangle,nakatsuji2023multiscale,foo2023extended,popov2023magic_butterfly} and experimental findings~\cite{uri2023superconductivity} have focused on the helical configuration of TTG~\cite{Christophe_2019,PhysRevLett.125.116404}. This specific structure is of interest due to the presence of two non-commensurate moiré interference patterns, resulting in a moiré quasiperiodic crystal~\cite{uri2023superconductivity}. In the case of a small twist angle, the separation between length scales leads to the introduction of an effective moiré lattice, defined as the closest commensurate ratio~\cite{Yuncheng2023}, along with a small deviation that gives rise to the supermoiré lengthscale~\cite{Yuncheng2023,guerci2023chern,popov2023magic_butterfly}.
The slow supermoir\'e periodicity can be seen as a relative shift between the two different moir\'e scales and parametrized with a displacement vector~\cite{Yuncheng2023,guerci2023chern}. 


Theoretical studies on helical trilayer graphene~\cite{devakul2023magicangle,nakatsuji2023multiscale,foo2023extended} have revealed the significant impact of lattice relaxation effects on the structure at the supermoiré scale. These studies have shown that energetically favorable ABA and BAB stacking regions expand at the expense of AAA regions. This leads to a real-space pattern characterized by triangular domains, where a sizeable gap separates the two central low-energy bands from the remote ones.
The ABA regions are characterized by a total Chern number $C_{\rm tot}=1$ for the central bands, while the BAB regions have $C_{\rm tot}=-1$, resulting in a real-space Chern mosaic~\cite{guerci2023chern}. This mosaic is separated by a network of chiral gapless regions.
In the chiral limit~\cite{Grisha_TBG}, the electronic bandstructure exhibits perfectly flat bands at a discrete series of magic angles. These flat bands possess interesting properties distinct from those observed in twisted bilayer graphene~\cite{Grisha_TBG,Grisha_TBG2,popov2020hidden,Ledwith_ann_2021,Wang_2021,Sheffer_2021,hFL_Bernevig_2022}.

The study of flat bands, or zero modes, in the chiral limit consists in finding the kernel of an operator with a purely holomorphic derivative and an abelian or non-abelian periodic potential~\cite{atiyah1984dirac,Non_Abelian_Graphene,PhysRevB.77.205424,Ledwith_ann_2021,Sheffer_2021,ledwith2022vortexability,gao2022untwisting,parhizkar2023generic}. Investigating the mathematical structure of this operator for helical trilayer graphene, Popov and Tarnopolsky~\cite{popov2023magic_butterfly} have recently identified two possible scenarios of flat bands for symmetric stackings (AAA, ABA, or BAB).
The first scenario, originally discussed in Ref.~\cite{guerci2023chern} (see also Ref.~\cite{devakul2023magicangle}), consists of two flat bands: a color-entangled Chern 2 band and a Chern -1 band. The color-entangled band~\cite{Barkeshli_2012,Wu_2013,YangLe_2014,Ledwith_Khalaf_2022,wang2022origin,dong2022,mera2023uniqueness,estienne2023ideal} is particularly interesting as it cannot be simply reduced to a single Landau level.
The second scenario, originally proposed in Ref.~\cite{popov2023magic}, features a fourfold degenerate flatband manifold with an additional Dirac cone crossing the flat bands at the $\Gamma$ point.


\begin{figure}
    \centering    \includegraphics[width=1.\linewidth]{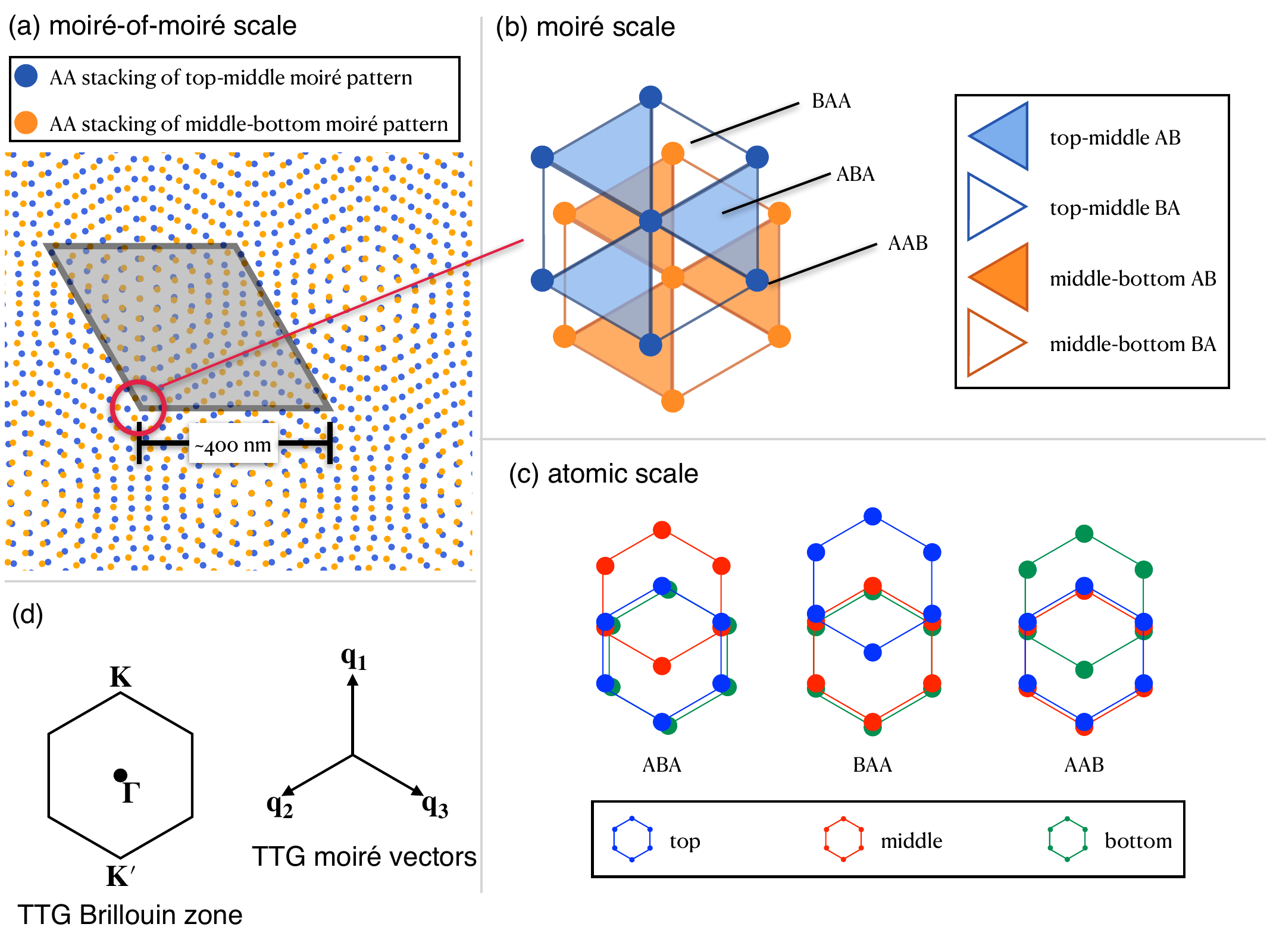}
    \caption{ (a) Long wavelength periodicity at the supermoir\'e lattice. Each blue and orange point corresponds to a position where a precise AA stacking occurs between, respectively, the top-middle and the middle-bottom layers. The blue and orange sets of points show the two incommensurate moiré patterns. (b) ABA configuration characterized by a relative shift of $\br_0=(\ba_1-\ba_2)/3$ between the two moir\'e patterns. (c) The resulting moir\'e pattern involves the atomic scale configurations ABA, BAA and AAB where the atom of one layer lie at the center of two overlying hexagons. (d) Momentum space Brillouin zone with Dirac zero modes at $K$, $K'$ and $\Gamma$, $\bq_j$ shows the vectors defining the periodicity at the moir\'e scale. }
\label{fig:sketch}
\end{figure}

In this work, we demonstrate that helical trilayer graphene with ABA (or BAB) stackings and equal twist angles displays a systematic series of magic angles, where the two scenarios are alternatively realized. Odd magic angles feature twofold degenerate flat bands following the first scenario, while even magic angles have four degenerate flat bands and a Dirac cone as per the second scenario. In each case, we prove the emergence of the flat band structure and derive analytical expression for the zero-energy wavefunctions and the resulting Chern numbers. 

Our theory reveals interesting relations between the dimension of the vector space spanned by the zero-energy modes and the total Chern number of the band. It provides an orthogonality relation between chiral and anti-chiral zero modes and demonstrates how the $C_{3z}$ and $C_{2y}T$ symmetries constrain the wavefunctions asymptotics and ultimately protect the emergence of the flat bands, in contrast with particle-hole symmetry which proves inessential. Finally, we briefly investigate the fate of the flat bands when breaking both the particle-hole and chiral symmetries.

The plan of the paper is as follows. In Sec.~\ref{sec_1} we introduce the model and the zero-mode equation in the chiral limit. Then, we discuss the symmetries of the model protecting three zero-energy Dirac cones at $K$, $K'$ and $\Gamma$ of the mini Brillouin zone in Sec.~\ref{subsec:symmetries}. Finally, we give in Sec.~\ref{geometrical} the geometrical relations connecting the chiral and anti-chiral flatband sectors. 
In Sec.~\ref{sec_2} we move to the investigation of the zero-modes discussing the different properties realized at even and odd magic angles. We provide the analytical solution of the wavefunctions of the different flatbands and we explain the different properties of odd Sec.~\ref{subsec:odd_magic} and even Sec.~\ref{subsec:even_magic} magic angles employing symmetry arguments as well as analytical results. In Sec.~\ref{sec:ph_breaking} we discuss the effect of the breaking of the particle-hole symmetry $P$.
Finally, we give a summary of the main results of our work in Sec.~\ref{sec:conclusions}.

\section{The chiral model for helical trilayer graphene}
\label{sec_1}

Helical trilayer graphene is formed by stacking three graphene layers in a staircase configuration with equal twist angles, as depicted in Fig.~\ref{fig:sketch}.
It shows two incommensurate moiré interference patterns, resulting in a superimposed modulation of the relative spatial shift between the two patterns. This long distance periodic modulation defines a triangular supermoiré, or moiré-of-moiré lattice. It can be seen as a slow variation of the atomic registry, first at the moiré scale and then at the supermoiré scale as shown in Fig.~\ref{fig:sketch}. 
The details of the model~\cite{guerci2023chern} we use for the single-particle Hamiltonian are given in Appendix~\ref{app:local_Hamiltonian}. It is a continuum model~\cite{Bistritzer12233,Santos,Santos2,Mele_2010} parametrized by the local relative shift between the two moirés. The parametrization evolves continuously between AAA, ABA and BAB local stacking configurations over the supermoiré unit cell.


The atomic relaxation in helical trilayer graphene has been investigated in recent works~\cite{devakul2023magicangle,nakatsuji2023multiscale,foo2023extended} where rearrangements of the atomic registry was demonstrated.
Theoretical calculations~\cite{devakul2023magicangle} (see also~\cite{nakatsuji2023multiscale,foo2023extended}) performed at small twist $\approx1.5^\circ$ have revealed that relaxation favours the formation of a triangular lattice with linear size $\approx400$nm of large domains characterized by the energetically favourable ABA/BAB stacking expanding at the expense of AAA regions, separated by a network of domain walls hosting chiral edge modes~\cite{nakatsuji2023multiscale}. 

In this work, we shall focus on the spatially hegemonic ABA stacking shown in Fig.~\ref{fig:sketch}b composed by the periodic modulation of the atomic configurations in Fig.~\ref{fig:sketch}c. The BAB stacking is simply the $C_{2z}T$ symmetric of ABA. 
Furthermore, we aim for simplicity and a deeper analytical understanding by considering the chiral limit~\cite{Grisha_TBG,Vafek_2020,Ledwith_ann_2021,Bultinck_2020_PRX,Khalaf_2021,bernevig2020tbg} of our continuum model, see Appendix~\ref{app:local_Hamiltonian}. Away from the chiral limit, the flat bands become dispersive but their bandwidth is still much smaller than the energy gap to the remote bands~\cite{guerci2023chern,devakul2023magicangle}, and they keep their topological features.


\subsection{Zero modes}

As discussed in Appendix~\ref{app:local_Hamiltonian}, the Hamiltonian for ABA stacking in the chiral limit takes the form
\begin{equation}
\label{H_ABA}
\mathcal H_{\rm ABA}(\br)=  \begin{pmatrix}
    0 & \mathcal D(\br) \\
    \mathcal D^\dagger(\br) & 0
\end{pmatrix} , 
\end{equation}
for a single graphene valley ($K$),
in the "sublattice-Chern" basis $(\psi_1,\psi_2,\psi_3,\chi_1,\chi_2,\chi_3)$~\cite{Ledwith_ann_2021,Bultinck_2020_PRX,Khalaf_2021,bernevig2020tbg}, where $\psi_\ell$ and $\chi_\ell$ refer respectively to the $A$ and $B$ sublattice, and $\ell=1,2,3$ labels the three distinct layers from top to bottom. We have introduced the differential operator
\begin{equation}
\label{matrix_operators}
   \mathcal D(\br) =-i\sqrt{2}I_{3\times 3}\partial + \mathcal A(\br), 
\end{equation}
with the derivative $\partial=(\partial_x-i\partial_y)/\sqrt{2}$ associated to the complex coordinate $z=(x+iy)/\sqrt{2}$,
and the non-abelian traceless gauge potential
\begin{equation}
\label{gauge_potential}
    \mathcal A(\br) =   
   \begin{pmatrix}
    0 & \alpha U_{\omega}(\br) & 0 \\
    \alpha U_{0}(-\br) & 0 & \alpha U_{0}(\br)  \\
    0 & \alpha U_{\omega}(-\br) & 0 
\end{pmatrix},
\end{equation}
resulting from electron tunneling between the layers with the dimensionless strength $\alpha$. Given the wavevector modulation $\bq_{j+1}= i e^{2i\pi j/3}$ shown in Fig.~\ref{fig:sketch}c, for which we use a complex notation~\cite{Christophe_2019}, 
the moiré potentials are given by
\begin{equation}
    U_0(\br) =  \sum_{j=1}^3 e^{-i\bq_j\cdot\br},
\end{equation}
$U_{\omega}(\br)=U_0(\br+\br_0)$ and $U_{\omega^*}(\br)=U_0(\br-\br_0)$ with $\br_0=(\ba_1-\ba_2)/3$. The moiré lattice vectors are $\ba_{j}=4\pi e^{i\pi/6}e^{i2\pi(j-1)/3}/3$ with $j=1,2$.
In the expressions above, all energy scales are expressed in unit of $v_F k_\theta$, and all momentum (length) scales in units of $k_\theta$ ($1/k_\theta$) with the moiré momentum $k_\theta=\theta K_D$, $K_D=4\pi/3a_{\rm G}$ and graphene lattice constant $a_{\rm G}\approx2.46$\AA. $\theta$ is the twist angle between consecutive layers and $\alpha=w_{\rm AB}/v_{F}k_\theta$.


An inspiring mathematical structure emerges~\cite{Grisha_TBG,guerci2023chern} in the chiral limit.  $\mathcal H_{\rm ABA}$ hence anticommutes $\{\mathcal H_{\rm ABA},\Lambda^z\}=0$ with the chiral operator
\begin{equation}
\label{chiral_operator}
 \Lambda^z =  \begin{pmatrix}
     I_{3\times3} & 0  \\ 
     0 & -I_{3\times 3} 
 \end{pmatrix},
\end{equation}
and the search for zero-energy modes decomposes into
\begin{equation}
\label{zero_mode_equation}
    \mathcal D(\br) \bm\chi_{\bk} (\br) = 0 ,\quad \mathcal D^\dagger(\br) \bm\psi_{\bk}(\br) =0, 
\end{equation}
where $\bm\psi$ and $\bm\chi$ are eigenstates of the chiral operator $\Lambda^z$ with positive $\Lambda^z=1$ (chiral sector) and negative $\Lambda^z=-1$  (anti-chiral sector) eigenvalue, respectively. The chiral and anti-chiral sectors also refer to the $A$-sublattice and $B$-sublattice polarized states, respectively.
The zero modes must also satisfy the Bloch periodic boundary conditions
\begin{equation}\label{bloch-periodic}
\begin{split}
    &\bm\psi_{\bk}(\br +\bm a_{1/2})=e^{i\bk\cdot\bm a_{1/2}}U_\varphi \bm\psi_{\bk}(\br),\\
&\bm\chi_{\bk}(\br +\bm a_{1/2})=e^{i\bk\cdot\bm a_{1/2}}U_\varphi \bm\chi_{\bk}(\br),
\end{split}
\end{equation}
inherited from Eq.~\ref{H_boundary_conditions}, with $U_{\varphi}=\diag[\omega^*,1,\omega]$.
The remainder of this paper will be devoted to analyzing the properties of the zero modes, which are solutions of Eq.\eqref{zero_mode_equation} under the boundary conditions specified in Eq.\eqref{bloch-periodic}.

\subsection{Symmetries and Dirac cones}
\label{subsec:symmetries}

The Hamiltonian $\mathcal H_{\rm ABA}$~\eqref{H_ABA} is invariant under the spatial symmetries $C_{3z}$ and $C_{2y}T$ forming the space group $P32'1$ ($\#150.27$ in the BNS notation~\cite{Aroyo2006_I,Aroyo2006_II}). In the sublattice basis we have: 
\begin{equation}
    C_{3z}\mathcal H_{\rm ABA}(\br) C^{-1}_{3z} = \mathcal H_{\rm ABA}(C_{3z}\br),
\end{equation}
where
\begin{equation}
\label{c3z}
    C_{3z} = \begin{pmatrix}
        \omega^* & 0 & 0 \\
        0 & 1 & 0 \\
        0 & 0 & \omega^*
    \end{pmatrix}\otimes\begin{pmatrix}
        \omega & 0 \\
        0 & \omega^*
    \end{pmatrix}.
\end{equation}
$C_{2y}T$ is the composition of a spinless time-reversal symmetry and a two-fold rotation around the $y$ axis: 
\begin{equation}
    C_{2y}T\mathcal H_{\rm ABA}(\br) \left(C_{2y}T\right)^{-1} = \mathcal H_{\rm ABA}(C_{2y}\br),
\end{equation}
where the transformation exchanges top and bottom layer and includes the complex conjugation operator $\mathcal K$:
\begin{equation}
\label{c2yT}
    C_{2y}T =\begin{pmatrix}
        0 & 0 & 1 \\
        0 & 1 & 0 \\
        1 & 0 & 0
    \end{pmatrix}\otimes I_{2\times 2}\mathcal K.
\end{equation}
In addition to these spatial symmetries, the model also exhibits an emerging particle-hole symmetry~\cite{Christophe_2019,Yuncheng2023,guerci2023chern}
\begin{equation}
\label{particle_hole}
    P= \begin{pmatrix}
        0 & 0 & 1 \\
        0 & -1 & 0 \\
        1 & 0 & 0
    \end{pmatrix}\otimes \sigma^0,
\end{equation}
anticommuting with $\mathcal H_{\rm ABA}$
\begin{equation}
    P \mathcal H_{\rm ABA}(\br)  P^{-1}= - \mathcal H_{\rm ABA}(-\br).
\end{equation}
In general, {\it i.e.} for all twist angles $\theta$, the Hamiltonian ${\cal H}_{\rm ABA}$  possesses three pairs (chiral and anti-chiral) of zero modes from which three Dirac cones emerge. They are located at  $\Gamma$, $K$ and $K'$ of the mini Brillouin zone Fid.~\ref{fig:sketch}c corresponding to $\bk = 0,\bq_1,-\bq_1$, and originate from the Dirac cones of the individual three graphene layers. The zero modes at $\Gamma$ are protected by particle-hole symmetry $P$ whereas those at $K$ and $K'$ are stabilized by the anti-unitary particle-hole operator $PC_{2y}T$, as further discussed in Appendix~\ref{subsec:ph_protection}. We remark that our symmetry analysis is also valid away from the chiral limit~\cite{Christophe_2019,Yuncheng2023}.

Magic angles are specific values of the twist angle $\theta$ (or $\alpha$ as they are related to each other) at which the two central bands of ${\cal H}_{\rm ABA}$ become perfectly flat. At these angles, the zero modes at $\Gamma$, $K$, and $K'$ are no longer unique, and an extensive degenerate set of zero modes emerges, forming flat bands.

\subsection{Geometrical relations}\label{geometrical}

Irrespective of the twist angle and value of $\alpha$, we can derive a set of identities which shows an interesting structure for the zero modes.
Chiral and anti-chiral zero-energy modes satisfy the relation: 
\begin{equation}
\label{orthogonality_relation}
v(\br) = \bar{\bm \chi}_{\bk_1}(\br)\cdot\bm\psi_{\bk_2}(\br)=0,    
\end{equation}
for $\bk_1\neq\bk_2$ and arbitrary $\br$ with $\bar{\bm \chi}_{\bk_2}(\br)\equiv {\bm\chi}^*_{\bk_2}(\br)$. 
The proof of Eq.~\eqref{orthogonality_relation} is straightforward. A direct computation shows that $\bar\partial v(\br)=0\implies v(\br)=v(z)$ by simply using that ${\bm \chi}_{\bk_1}(\br)$ and $\bm\psi_{\bk_2}(\br)$ are zero modes of $\mathcal D(\br)$ and $\mathcal D^\dagger (\br)$, respectively. $v(z)=0$ then follows from Liouville's theorem and periodicity over the moir\'e unit cell. Eq.~\eqref{orthogonality_relation} tells us that the chiral $\bm\psi$ and anti-chiral $\bar{\bm\chi}$ solutions of different momenta at a given $\br$ are orthogonal to each other.
In addition, we also find that a chiral zero mode can be generated from a pair of anti-chiral solutions. Namely,
\begin{equation}
\label{generate_chiral}
    \bm\psi_{-\bk_1-\bk_2}(\br)
    = \bar{\bm\chi}_{\bk_1}(\br)\times\bar{\bm\chi}_{\bk_2}(\br)
\end{equation}
solves $\mathcal D^\dagger(\br) \bm\psi_{-\bk_1-\bk_2}(\br)=0$. Similarly, 
\begin{equation}
\label{generate_antichiral}
 \bm\chi_{-\bk_1-\bk_2}(\br)   = \bar{\bm\psi}_{\bk_1}(\br)\times\bar{\bm\psi}_{\bk_2}(\br),
\end{equation}
solves $\mathcal D(\br){\bm\chi}_{-\bk_1-\bk_2}(\br)=0$. Expressions similar to Eq.~\eqref{orthogonality_relation} and Eq.~\eqref{generate_chiral} have been derived in Ref.~\cite{popov2023magic_butterfly} but with a different choice of gauge.

\begin{figure}
    \centering    \includegraphics[width=0.9\linewidth]{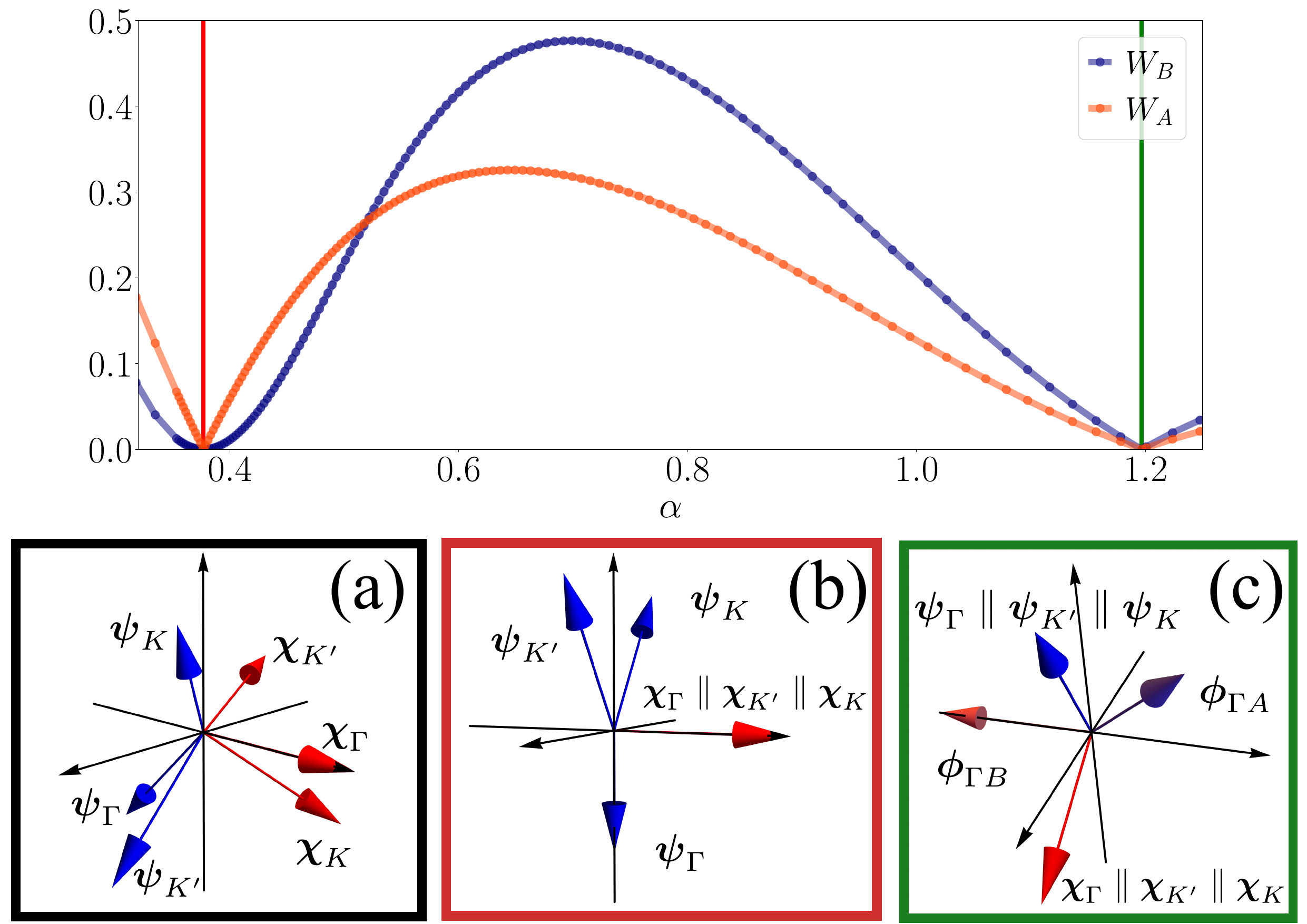}
    \caption{Top panel: Wronskian $|W_A|$ (orange) and $|W_B|$ (blue) for the chiral and anti-chiral sector as a function of $\alpha$. $|W_{A/B}|$ vanishes at the magic angles denoted as red and green vertical lines. Bottom panel: Figs.~(a)-(c) show a schematic representation of the zero mode spinors $\{K,K',\Gamma\}$ in the chiral and anti-chiral sectors. Panel (a) represents a generic configuration away from the magic angle where the spinors are linearly independent and chiral and anti-chiral sectors satisfy Eq.~\eqref{orthogonality_relation}, while (b) and (c) sketch the two scenario realized at the magic angle in the helical trilayer graphene. In panel (c) $\bm\phi_{\Gamma A}$ and $\bar{\bm\phi}_{\Gamma B}$ are the two additional zero modes degenerate with the flat bands at $\Gamma$.    }
\label{fig:wronskian}
\end{figure}

We introduce the Wronskian~\cite{guerci2023chern,popov2023magic_butterfly} of the Dirac spinors in the chiral sector:
\begin{equation}
\label{wronskian_A}
    W_A(\br) = \bm\psi_{\Gamma}(\br)\cdot\left[\bm\psi_{K}(\br)\times\bm\psi_{K'}(\br)\right],
\end{equation}
which obeys the relation $\bar\partial W_A=0$. 
Thus, according to Liouville's theorem, we have $W_A(\mathbf{r}) = W_A$. $W_A\neq0$ since, unlike $v(\mathbf{r})$, Eq.~\eqref{wronskian_A} is invariant under translations of moiré lattice vectors. $W_A\neq0$ implies that the three vectors $\bm\psi_\Gamma$, $\bm\psi_K$, and $\bm\psi_{K'}$ are linearly independent.
Similarly, we can define the Wronskian in the anti-chiral sector
\begin{equation}
\label{wronskian_B}
    W_B(\br)=\bm\chi_{\Gamma}(\br)\cdot\left[\bm\chi_{K}(\br)\times\bm\chi_{K'}(\br)\right]
\end{equation}
 satisfying the condition $\partial W_B=0$, which implies $W_B(\br)=W_B$. Combining the definitions of the Wronskians with Eqs.\eqref{generate_chiral} and\eqref{generate_antichiral}, we find that $W_A$ and $W_B$ are both proportional, up to a phase, to the scalar products
 \begin{equation}\label{scalar}
   \bm\bar{\chi}_{\Gamma}(\br) \cdot \bm\psi_{\Gamma}(\br), \quad  \bm\bar{\chi}_{K}(\br) \cdot \bm\psi_{K}(\br), \quad \bm\bar{\chi}_{K'}(\br) \cdot \bm\psi_{K'}(\br),
 \end{equation}
  between the chiral and anti-chiral zero modes at $\Gamma$, $K$, and $K'$. Consequently, $W_A$ and $W_B$ vanish simultaneously with these scalar products. As illustrated in the top panel of Fig.~\ref{fig:wronskian}, the simultaneous vanishing of  $|W_A|$ and $|W_B|$ defines the positions of the magic angles~\cite{guerci2023chern,popov2023magic_butterfly}.

From these different relations, an intuitive picture emerges. At non-magic angles, the set of  chiral vectors ${\cal C}_A \equiv \{\bm\psi_\Gamma,\bm\psi_K,\bm\psi_{K'}\}$ is linearly independent and span the full three-dimensional space of layers at each $\br$. The same holds for the anti-chiral set ${\cal C}_B \equiv \{\bm\chi_\Gamma,\bm\chi_K,\bm\chi_{K'}\}$ whereas chiral and anti-chiral vectors are mutually orthogonal when their momenta are different as a result of Eq.~\eqref{orthogonality_relation}. This is depicted in Fig.~\ref{fig:wronskian}a. The situation is quite different at magic angles because the scalar products of Eq.~\eqref{scalar} all vanish. The chiral ${\cal C}_A$ and anti-chiral ${\cal C}_B$ sets then form two distinct subspaces which are orthogonal to each other with the three-dimensional layer space. There are three possible cases, illustrated in Fig.~\ref{fig:wronskian}b and c - corresponding to the two scenarios of Ref.~\cite{popov2023magic_butterfly} - (i) the chiral set forms a subspace of dimension 2 and the anti-chiral set is forced to be one-dimensional, {\it i.e.} all three anti-chiral vectors are collinear. (ii) same as (i) but the role of chiral and anti-chiral are exchanged. (iii) both chiral and anti-chiral sets have dimension 1. 

As we will demonstrate in Sec.~\ref{sec_2}, the chiral and anti-chiral flat bands that emerge at the magic angle are generated by the two sets ${\cal C}_A$ and ${\cal C}_B$, leaving the subspace decomposition invariant: $2+1$ for (i), $1+2$ for (ii), and $1+1$ for (iii). Additionally, for case (iii), we will show that the third dimension is filled by a pair of additional zero modes at the $\Gamma$ point, corresponding to a crossing of the flat bands by a Dirac cone.


By decreasing the twist angle or increasing the value of $\alpha$, we will uncover in Sec.~\ref{sec_2} a series of magic angles alternating between cases (i) and (iii) with an even/odd effect. Interestingly, we will also observe that the dimension or rank of the flat band coincides with the absolute value of the Chern number, highlighting the relationship between the rank of the flat band and the number of lowest Landau levels comprising it~\cite{wang2022origin,dong2022,mera2023uniqueness,estienne2023ideal}.

\begin{figure}
    \centering    \includegraphics[width=0.95\linewidth]{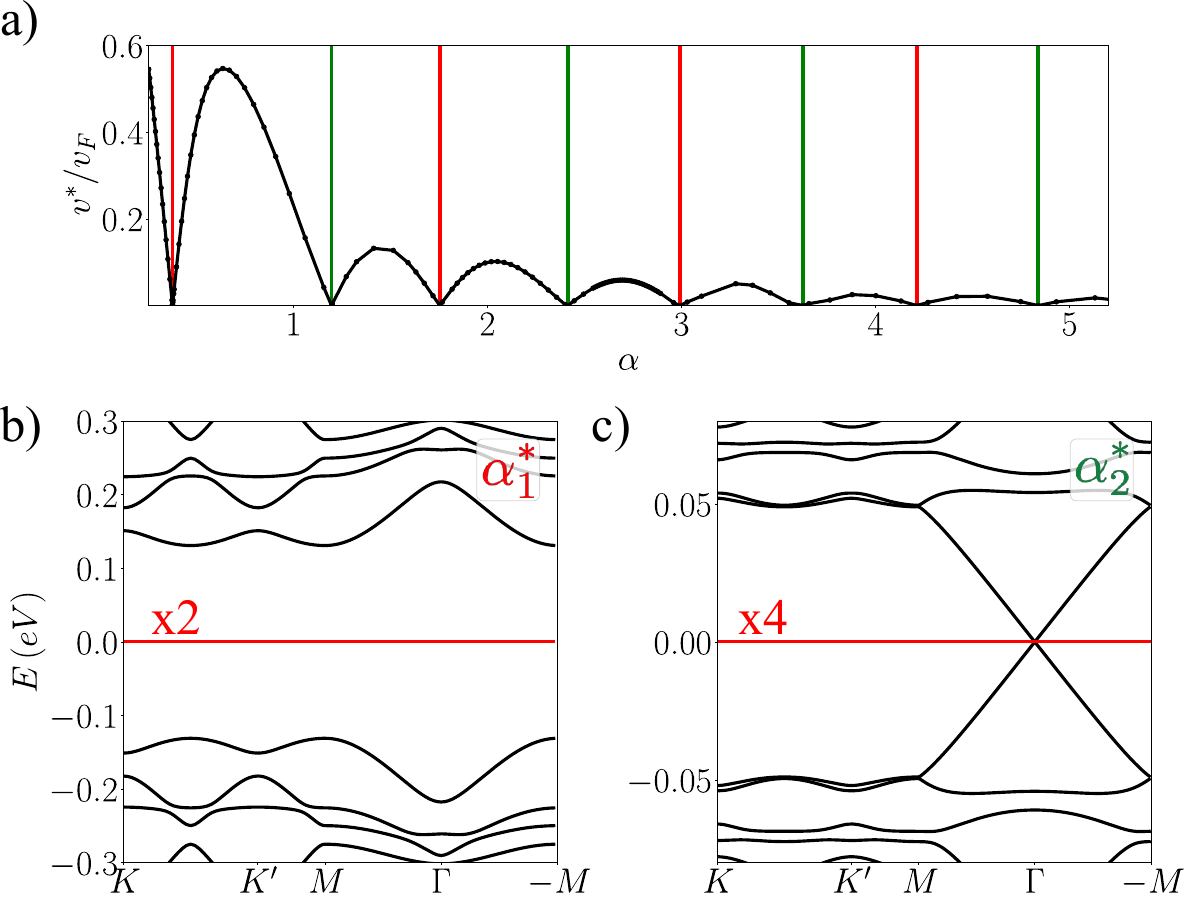}
    \caption{ a) Renormalized Fermi velocity at $K$ as a function of $\alpha$. Vertical red and green lines denote odd $\alpha^*_{2n-1}$ and even $\alpha^*_{2n}$ magic angles, respectively. Panel b) and c) show the bandstructure at the first magic angle $\alpha^*_1\approx0.377$ ($\theta^*_1\approx 1.687^\circ$) and the second magic angle $\alpha^*_2\approx1.197$ ($\theta^*_2\approx 0.532^\circ$), respectively. Red lines denote the flat bands, while $\times 2$ and $\times 4$ give the number of zero modes per $\bk$-point for odd and even magic angles, respectively.}
\label{fig:flatbands_1st_2nd}
\end{figure}

\section{Even and odd series of magic angles}
\label{sec_2}

The sequence of consecutive magic angles is showcased by computing the renormalized Fermi velocity at $K$ as a function of $\alpha$, see Fig.~\ref{fig:flatbands_1st_2nd}a, in agreement with the Wronskians shown in  Fig.~\ref{fig:wronskian}. The odd and even magic angles are highlighted by red and green vertical lines and correspond, respectively, to the cases (i) and (iii) discussed in Sec.~\ref{geometrical}. For odd magic angles $\alpha^*_{2n-1}$, the first being $\alpha^*_1\approx0.377$ ($\theta^*_1\approx1.687^\circ$), the spectrum features two degenerate flat bands as displayed in Fig.~\ref{fig:flatbands_1st_2nd}b. For even magic angles $\alpha^*_{2n}$ however, the first one is $\alpha^*_2\approx1.197$ ($\theta^*_2\approx0.532^\circ$), four degenerate flat bands arise coexisting with a Dirac cone crossing them at $\Gamma$, as shown in Fig.~\ref{fig:flatbands_1st_2nd}c. The structure and degeneracy of the zero-energy bands thus repeats periodically and depends on the parity of the magic angle label. 

\begin{figure}
    \centering
    \includegraphics[width=0.95\linewidth]{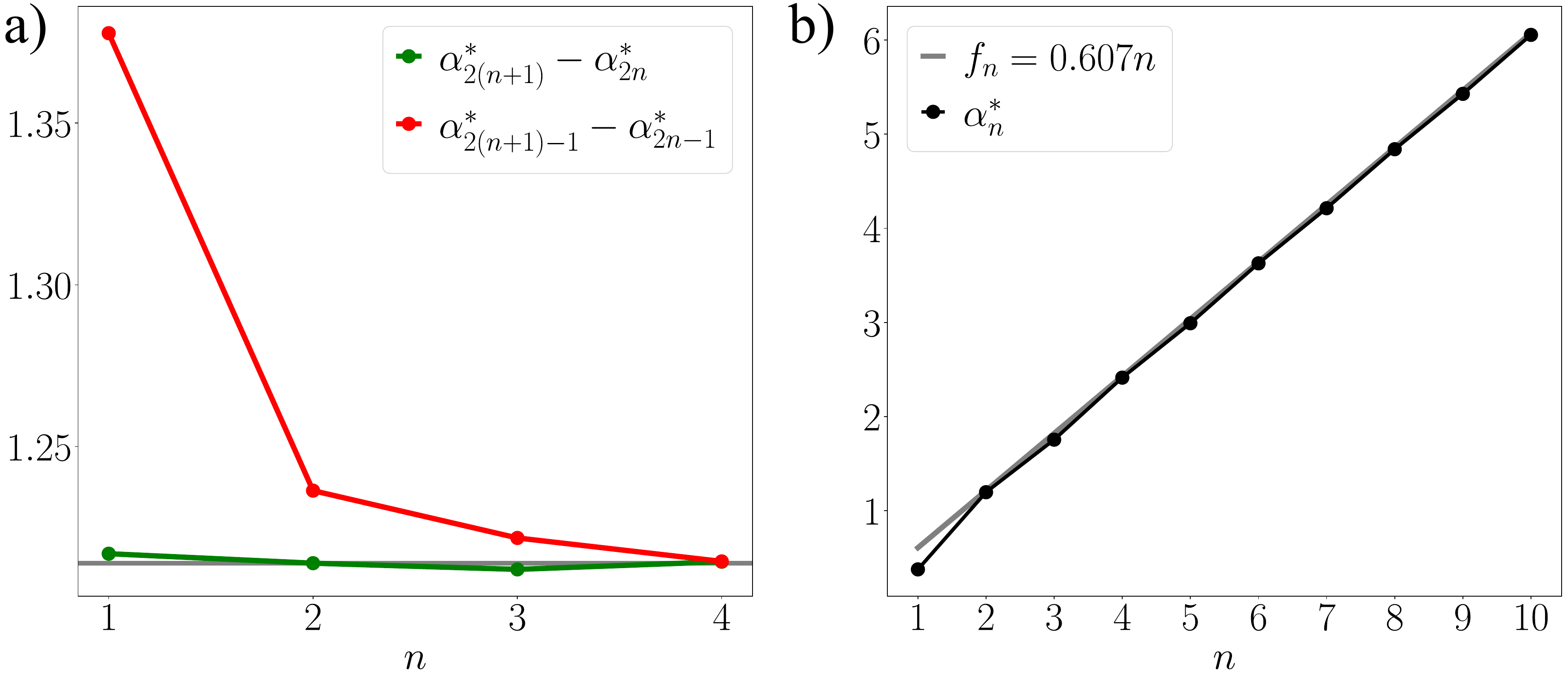}
    \caption{a) Distance between neighboring magic angles in the even and odd sequence. b) Sequence of magic angles $\alpha_n$. Increasing the order $n$ the distance between nearest neighbour magic angles in the even and odd sectors approaches a constant value $\approx 1.214$ represented by the horizontal gray line in a). }
    \label{fig:magic_angle_sequence}
\end{figure}

Furthermore, inspecting more closely the sequence of magic angles, we find that the difference between consecutive values rapidly approaches a constant value $\alpha^*_{2n+1}-\alpha^*_{2n-1}\approx \alpha^*_{2n+2}-\alpha^*_{2n}\simeq 1.214$, as shown in Fig.~\ref{fig:magic_angle_sequence}, similarly to the twisted bilayer case~\cite{Grisha_TBG,RafeiRen_TBG,Becker_2022,Naumis_2023}.
In the following, we demonstrate that the distinct nature of the even and odd magic angles arises from symmetry considerations. These constraints dictate the behavior of the zero-mode wavefunction at the high-symmetry points together with the identities presented in Sec.~\ref{geometrical}.

\subsection{Zero modes for odd magic angles}
\label{subsec:odd_magic}

\begin{figure}
    \centering
    \includegraphics[width=0.95\linewidth]{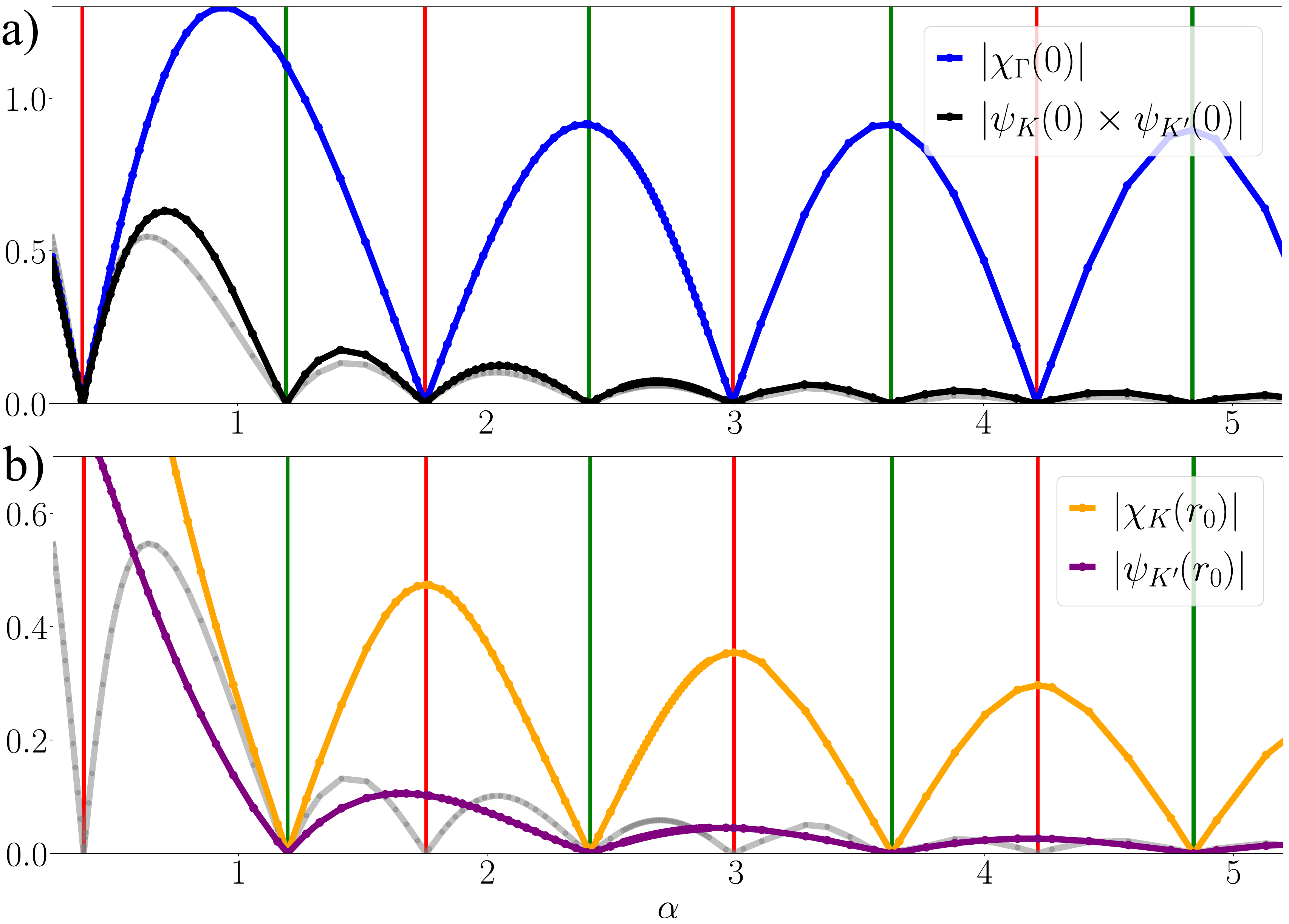}
    \caption{a) Wavefunction absolute value $|\bm \chi_\Gamma(0)|$ (blue) and cross product $|\bm\psi_K(0)\times\bm\psi_{K'}(0)|$ (black) as a function of $\alpha$. b) Wavefunction absolute values $|\bm \chi_K(\br_0)|$ and $|\bm \psi_{K'}(\br_0)|$ as a function of $\alpha$. The grey solid line shows the renormalized Fermi velocity $v^*/v_F$. Vertical green and red lines show the location of odd and even magic angle, respectively. }
\label{fig:magic_angle_diagnostic}
\end{figure}

To gain analytical insight into the origin of the odd magic angle, we study the anti-chiral zero mode $\bm\chi_\Gamma(\br)$ in the vicinity of the high-symmetry point AA ($\br=0$).
The wavefunctions around AA regions are not described by pseudo Landau levels~\cite{XiDai_PseudoLandaulevel}, reflecting a charge distribution of the flatbands different from the one realized in twisted bilayer graphene~\cite{devakul2023magicangle}.
As detailed in Appendix~\ref{app:odd_magic}, the wavefunction $\bm\chi_\Gamma(\br)$ can be formally expanded in powers of $z$ and $\bar{z} \equiv z^*$ close to $r=0$. Enforcing the symmetries $C_{3z}$, $C_{2y}T$ and $P$ constrains and simplifies the resulting expansion. To leading order, we find
\begin{equation}
\label{B_AA}
 \bm  \chi_\Gamma(\br)= \begin{pmatrix}
     0  \\
     \chi_{ 2}  \\
     0 
 \end{pmatrix} +  O(\bar z),
\end{equation}
where $C_{2y}T$ (see Eq.~\eqref{c2yT}) imposes that $\chi_2\equiv\chi_{\Gamma 2}(0)$ is a real coefficient. Plotting the real $\chi_2$ as a function of $\alpha$ in Fig.~\ref{fig:magic_angle_diagnostic}a, we find that it vanishes for all odd magic angles. The condition $\chi_2=0$, implying $\bm  \chi_\Gamma(0)=0$, thus defines the series of odd magic angles and yields a vanishing Wronskian from Eq.~\eqref{wronskian_B}. The vertical red lines of Fig.~\ref{fig:magic_angle_diagnostic} exactly matches the one obtained from the Wronskian in Fig.~\ref{fig:wronskian}.

Beyond Eq.~\eqref{B_AA}, the next order obeying symmetries is 
$\bm\chi_{\Gamma}(\br) \simeq \bar z (\chi'_1,0,{\chi'}^*_1)$ at the odd magic angles. Here, $\chi'_1$ is an arbitrary complex coefficient dependent on the magic angle. The presence of this simple zero is sufficient to predict and explicitly construct~\cite{guerci2023chern} the whole anti-chiral flat band following the seminal reasoning of Tarnopolsky et al.~\cite{Grisha_TBG} in twisted bilayer graphene, see also Refs.~\cite{Grisha_TBG2,Ledwith_ann_2021,Wang_2021}. 
The zero-mode wavefunctions exhibit the analytical expression: 
\begin{equation}
\label{Chern_1_B_odd}
    \bm\chi_{\bk}(\br) = \bar\eta_{\bk}(\bar z)\bm\chi_\Gamma(\br),
\end{equation}
where the antiholomorphic $\bar\eta_{\bk}(\bar z)=\eta^*_{\bk}(-z)$ is related to the meromorphic function describing the lowest Landau level (LLL) on a torus~\cite{haldanetorus1,Haldane_2018}: 
\begin{equation}\label{holomorphic}
   \eta_{\bk} (z) = e^{i k_1 z/a_1} \frac{\vartheta_1[ z/a_1 - k/b_2,\omega]}{\vartheta_1[ z/a_1,\omega]}
\end{equation}
with the notation $k_1=\bk\cdot\ba_1$ and the Jacobi theta-function
\begin{equation}
    \vartheta_1[ z , \omega] = \sum_{n \in \mathbb{Z} } e^{i \pi \omega (n+1/2)^2} e^{2 i \pi (z-1/2)(n+1/2)}
\end{equation}
which vanishes at $z=0$ and results in a Bloch periodicity~\eqref{bloch-periodic} for Eq.~\eqref{Chern_1_B_odd}.
Momentum space boundary conditions~\cite{Wang_2021} on the self-periodic part of the Bloch state $\bm u_{\bar k} (\br) = \bm\chi_{\bk}(\br) e^{- i \bk \cdot \br} $ give
\begin{equation}
\label{k_space_BC}
{\bm u}_{\bar k+\bar b_j}( \br)= e^{-i{\bm b}_j\cdot \br }e^{i\phi_{{ k, b_j}}} {\bm u}_{\bar k}(\br), 
\end{equation}
with $\phi_{{ k, b_1}}=-2\pi\bar k/\bar b_2+\pi-\pi \bar b_1/\bar b_2$ and $\phi_{{ k, b_2}}=\pi$ corresponding to a flat band with total Chern number $C_B=-1$~\cite{Wang_2021}. 

Thanks to the spatial symmetries, the vanishing of $\chi_2$ at the odd magic angles is thus sufficient to predict a flat band~\eqref{Chern_1_B_odd} in the anti-chiral sector with Chern number $-1$. Moreover, employing  Eq.~\eqref{generate_antichiral} also determines the chiral sector. Eq.~\eqref{generate_antichiral}, evaluated at the origin $\br = 0$ with $\bk_1=K$ and $\bk_2=K'$, gives 
\begin{equation}
\label{magic_relation}
  \bm \psi_{K} (0) \times  \bm \psi_{K'} (0) =\bar{\bm \chi}_\Gamma (0) =   0,
\end{equation}
resulting in the fact that ${\bm \psi}_{K} (0) $ and ${\bm \psi}_{K'} (0)$ are collinear vectors, see Fig.~\ref{fig:magic_angle_diagnostic}a. Expanding ${\bm \psi}_{K}$ and ${\bm \psi}_{K'}$ around $\br=0$ and enforcing the symmetries as in Appendix~\ref{app:odd_magic}, we find to leading order
$\bm\psi_K(0) = (\psi_1,0,\psi_3)$ and $\bm\psi_K(0) = (\psi_3^*,0,\psi_1^*)$ where $\psi_{1/3}$ are complex coefficients. Using Eq.~\eqref{magic_relation} yields
\begin{equation}
\label{magic_relation_1}
    \chi_2 = |\psi_3|^2-|\psi_1|^2=0\implies \psi_1=\psi_3 e^{i\varphi}.
\end{equation}
We choose a gauge with $\varphi=0$ and find
\begin{equation}
\label{magic_relation_2}
    \bm\psi_K(0)=\bm\psi_{K'}(0).
\end{equation}
As shown in Ref.~\cite{guerci2023chern}, this identity is sufficient to construct the flat band wavefunctions for the chiral sector,
\begin{equation}\label{flatAband}
   {\bm \psi}_{\bk}(\br) =   a_{k} 
   \eta_{\bk +K'} (z) {\bm \psi}_{K}(\br) +  a_{-k} \eta_{\bk + K} (z) {\bm \psi}_{K'}(\br),
\end{equation}
satisfying the Bloch periodicity Eq.~\eqref{bloch-periodic}, with the holomorphic function defined in Eq.~\eqref{holomorphic} and $a_{k} = \vartheta_1 [ (k+K)/b_2,\omega]$. Momentum space boundary conditions give $C_A=2$ as shown in Ref.~\cite{guerci2023chern}. The magic relation~\eqref{magic_relation_2} which is not associated to the vanishing of the chiral spinor $\bm\psi$ justifies the charge density distribution homogeneity in the Chern 2 band~\cite{devakul2023magicangle}. 

Remarkably, all odd magic angles feature a chiral flat band of Chern $2$, generated by the two vectors ${\bm \psi}_{K}(\br)$ and ${\bm \psi}_{K'}(\br)$, alongside an anti-chiral flat band of Chern $-1$ where all states align collinearly with $\bm\chi_\Gamma(\br)$. Moreover, Eq.~\eqref{orthogonality_relation} reveals that the chiral and anti-chiral flat band spaces are orthogonal to each other. Specifically, $\bm\chi_\Gamma(\br)$ and all the anti-chiral wavefunctions are oriented in a direction perpendicular to the chiral plane formed by ${\bm \psi}_{K}(\br)$ and ${\bm \psi}_{K'}(\br)$. This provides a clear understanding of why a chiral Chern $\pm 2$ band is  accompanied by an anti-chiral Chern $\mp 1$ band within a three-layer system.


Finally, we observe that the cross product $\bm\psi_{K}(0)\times\bm\psi_{K'}(0)$ vanishes also for even magic angles, see black line in Fig.~\ref{fig:magic_angle_diagnostic}a but for a different reason that we shall explain in the next Section.  


\subsection{Zero-modes for even magic angles}
\label{subsec:even_magic}

We now turn to the characterization of the zero-energy modes and flat bands for even magic angles. 
In contrast to the previous case, $\bm\chi_{\Gamma}(0)$ remains finite at $\alpha^*_{2n}$ as shown in Fig.\ref{fig:magic_angle_diagnostic}a. Consequently, the zero-mode construction discussed in Section\ref{subsec:odd_magic} does not apply for even magic angles.
To make progress, we examine the behavior of the zero-mode solution $\bm\psi_{K'}(\br)$ in the vicinity of the AB stacking point $\br_0=(\ba_1-\ba_2)/3$. By expanding the zero-mode equation~\eqref{zero_mode_equation} for the middle layer component $\psi_{K'2}$ to linear order in the deviation $\br$, we obtain:
\begin{equation}
  \label{psi_middle_layer}
    \bar\partial^2\psi_{K'2} - \frac{9\alpha^2}{2}\left(\frac{z^2}{2}+\frac{\bar z}{\sqrt{2}}\right)\psi_{K'2}=0,
\end{equation}
while top and bottom layer amplitudes are given by 
\begin{equation}
 \begin{split}
 &\bar\partial  \psi_{K'1}=3\alpha z\psi_{K'2}/\sqrt{2},\\
&\psi_{K'3}=\left(i\sqrt{2}\bar\partial \psi_{K'2}-3\alpha z\psi_{K'1}/\sqrt{2}\right)/(3\alpha). 
 \end{split}   
\end{equation}
We obtain the general solution of Eq.~\eqref{psi_middle_layer}: 
\begin{equation}
\label{psi_middle_layer_1}
    \psi_{K'2}(\br+\br_0) = \gamma_A Ai(\zeta) + \gamma_B Bi(\zeta), 
\end{equation}
where $\zeta = (2\alpha)^{2/3}\left(z^2/2+\bar z/\sqrt{2}\right)$ and we retain both Airy functions $Ai(z)$ and $Bi(z)$~\cite{NIST:DLMF}. A similar behavior of the zero-mode wavefunction has been obtained for twisted bilayer graphene around the AB/BA points in Ref.~\cite{RafeiRen_TBG}. $C_{3z}$ rotation centered at the high-symmetry point $\br_0$ implies: 
\begin{equation}
\label{C3z_r0}
    \bm\psi_{K'}(C_{3z}\br+\br_0)=\begin{pmatrix}
        \omega^* & 0 & 0 \\
        0 & \omega^* & 0 \\
        0 & 0 & 1 
    \end{pmatrix}\bm\psi_{K'}(\br+\br_0),
\end{equation}
employing the McLaurin expansion of the Airy functions~\cite{NIST:DLMF} we find $\gamma_A=-3^{2/3} N$, $\gamma_B=3^{1/6}N$:
\begin{equation}
\label{expansion_r0}
 \bm\psi_{K'}(\br+\br_0)=\begin{pmatrix}
     0 \\
     0\\
     \psi_3^{(0)}
 \end{pmatrix}-\frac{3i\alpha\psi_3^{(0)}}{\sqrt{2}}\begin{pmatrix}
     0 \\
     \bar z\\
     0
 \end{pmatrix}+O(z^2),
\end{equation}
where $\psi_3^{(0)}\equiv \psi_{K'3}(\br_0)$. 
In addition, the $PC_{2y}T$ symmetry gives: 
\begin{equation}
\label{PC2yT_r0}
    \bm\psi_{K'}(\br+\br_0)=\begin{pmatrix}
        -1 & 0 & 0 \\
        0 & 1 & 0 \\
        0 & 0 & -1 
    \end{pmatrix}\bm\psi^*_{K'}(C_{2x}\br+\br_0),
\end{equation}
implying $\Re\psi^{(0)}_3=0$. Consequently, $\bm\psi_{K'}(\br_0)$ is determined by a single real number $\Im\psi^{(0)}_3$. The condition for the even magic angle is, therefore, the vanishing of this coefficient, such that $\bm\psi_{K'}(\br_0)=0$, and the Wronskian $W_A$ also becomes zero as per Eq.\eqref{wronskian_A}. The vertical green lines displayed in Fig.~\ref{fig:magic_angle_diagnostic}b align with those in Fig.~\ref{fig:wronskian}.
At even magic angles, the wavefunction from Eq.~\eqref{expansion_r0} further expands as $\bm\psi_{K'}(\br_0+\br)\simeq z^2 (a,b,0)$ at small $\br$ (with complex coefficients $a$ and $b$), corresponding to a double zero.
This structure enables the following analytical form for the chiral flat bands
\begin{equation}
\label{analytical_solution_ABA_2}
    \bm\psi_{\bk}(\br) = \eta^{(0)}_{\bk'}(z)\eta^{(0)}_{\bk-\bk'+K}(z)\bm\psi_{K'}(\br),
\end{equation}
with the meromorphic and periodic function
\begin{equation}
    \eta^{(0)}_{\bk} (z) = e^{i k_1 z/a_1} \frac{\vartheta_1[ (z-z_0)/a_1 - k/b_2,\omega]}{\vartheta_1[ (z-z_0)/a_1,\omega]},
\end{equation}
$z_0$ denotes the complex representation of $\br_0$. 
Despite the fact that $\bk'$ is an arbitrary wavevector in Eq.~\eqref{analytical_solution_ABA_2}, at most two distinct values of $\bk'$ yield independent wavefunctions.
 Setting for instance, $\bk' = K'$ and $\bk' = 0$, we obtain two degenerate flat bands described by the wavefunctions $\bm\psi^{(1)}_{\bk}(\br) =\eta^{(0)}_{K'}(z)\eta^{(0)}_{K'+\bk}(z)\bm\psi_{K'}(\br)$ and $\bm\psi^{(2)}_{\bk}(\br) = \eta^{(0)}_{\bk+K}(z)\bm\psi_{K'}(\br)$. 

A very similar analytical form was introduced by Popov and Tarnopolsky~\cite{popov2023magic,popov2023magic_butterfly} to describe the fourfold degenerate flat bands but at AAA stacking magic angles. Mathematically, the construction of Eq.~\eqref{analytical_solution_ABA_2} is possible because the poles brought by the two functions $\eta^{(0)}_{\bk}$ at $z=z_0$ are precisely cancelled by the double zero of $\bm\psi_{K'} (\br)$ at $\br_0$ ($z_0$). The resulting wavefunctions are finite everywhere, obey the Bloch periodic boundary conditions Eq.~\eqref{bloch-periodic}, and solve the zero-mode Eq.~\eqref{zero_mode_equation}. All chiral flat band wavefunctions are collinear to $\bm\psi_{K'}(\br)$ and thus span a one-dimensional space. It readily explains the vanishing of the cross product $\bm\psi_{K}(0)\times\bm\psi_{K'}(0)$ also at even magic angle, as shown in Fig.~\ref{fig:magic_angle_diagnostic}a. The above derivation relies on the double zero of $\bm\psi_{K'}$  at $\br_0$. Alternatively, we can reconstruct the twofold degenerate flat band from the quadratic vanishing of $\bm\psi^{(1)}_{\Gamma}(\br) $ at $\br =0$. As discussed in Appendix~\ref{app:breaking-P}, the advantage of this latter approach is that it does not require the particle-hole symmetry $P$, only $C_{3z}$ and $C_{2y}T$.


Moving to the anti-chiral sector, we can repeat the same analysis for the wavefunction $\bm\chi_{K}(\br)$ in the vicinity of $\br_0$. $\bm\chi_{K}(\br_0)=0$ at even magic angles, as shown in Fig.~\ref{fig:magic_angle_diagnostic}b, and the expansion around $r_0$ is quadratic indicating a double zero. The same construction thus extends to the anti-chiral sector, resulting in two degenerate flat bands. Additionally, by applying momentum space boundary conditions, we can determine that each chiral flat band possesses a Chern number $C_A=+1$, whereas the anti-chiral flat bands have $C_B=-1$.
In total, we prove that even magic angles feature a two-fold degenerate set of bands in each sublattice sector, or four flat bands with a vanishing total Chern number.

The chiral flat bands align with $\bm\psi_{K'}(\br)$ while the anti-chiral with $\bm\chi_{K}(\br)$, leaving room for a third direction in the layer space. By following the arguments presented in Ref.\cite{popov2023magic_butterfly}, one can analytically demonstrate the existence of a pair of additional zero modes at $\Gamma$. These zero modes correspond to a Dirac cone crossing the flat bands in Fig.\ref{fig:flatbands_1st_2nd}c. The proof is outlined as follows.
From Eq.~\eqref{orthogonality_relation}, we know that $\bm\psi_{K}(\br)$ and $\bm\chi_{K'}(\br)$ are orthogonal to each other. Both wavefunctions vanish at $\br_0$ and remain finite everywhere else. Consequently, we can define a function $\bm\phi_\Gamma(\br)$ such that:
\begin{equation}
\label{newfunctionGamma}
    \bm\psi_{K'}(\br)=\bar{\bm\phi}_{\Gamma}(\br)\times\bar{\bm\chi}_{K}(\br),
\end{equation}
Here, we emphasize that this expression does not uniquely define $\bm\phi_\Gamma (\br)$. 
The subscript $\Gamma$ indicates that $\bm\phi_\Gamma (\br)$ has the Bloch periodicity of the $\Gamma$ point, as inferred from the periodicity of the two other functions. 
Applying the operator $\mathcal D^\dagger( \br)$ to both sides of Eq.~\eqref{newfunctionGamma}, we arrive at
\begin{equation}
    0 = - \left [ \mathcal D^*( \br) \bar{\bm\phi}_\Gamma (\br) \right] \times \bar{\bm\chi}_{K} (\br),
\end{equation}
which shows that $\mathcal D^*( \br) \bar{\bm\phi}_\Gamma (\br)$ must be proportional to $\bar{\bm\chi}_{K} (\br)$, or
\begin{equation}
   \mathcal D^*( \br) \bar{\bm\phi}_\Gamma (\br) = f(\br) \bar{\bm\chi}_{K} (\br)
\end{equation}
with some periodic function $f(\br)$. We introduce the function $g(\br)$, solution of $\bar{\partial} g(\br) = f(\br)$, and shift $\bar{\bm\phi}_\Gamma$ as
\begin{equation}
    \bar{\bm\phi}_\Gamma (\br) \to \bar{\bm\phi}_\Gamma (\br) - ig(\br) \bar{\bm\chi}_{K} (\br)/\sqrt{2}
\end{equation}
to finally obtain
\begin{equation}
   \mathcal D( \br) {\bm\phi}_\Gamma (\br) = 0.
\end{equation}
This last equation demonstrates that we have constructed an additional anti-chiral zero-energy solution. Due to its definition in Eq.\eqref{newfunctionGamma}, $\bm\phi_\Gamma(\br)$ cannot be proportional to $\bm\chi_{K}(\br)$ and therefore lies outside the one-dimensional anti-chiral subspace. Eq.\eqref{newfunctionGamma} also implies that $\bar{\bm\phi}_\Gamma(\br) \cdot \bm\psi_{K'}(\br)=0$: $\bm\phi_\Gamma(\br)$ and $\bm\psi_{K'}(\br)$ are orthogonal. As a result, $\bm\phi_\Gamma(\br)$ occupies the third vacant direction.
As already discussed earlier, $\bm\phi_\Gamma(\br)$ satisfies the Bloch periodic condition with momentum at $\Gamma$. A similar construction yields a second zero-energy state at $\Gamma$ in the chiral sector, also spanning the third direction orthogonal to both the chiral and anti-chiral flat bands. This result concludes the characterization of the low-energy spectrum at even magic angles.

\section{Breaking the particle-hole symmetry $P$}
\label{sec:ph_breaking}

The twist-angle dependency in the Pauli matrices obtained by replacing $\bm\sigma$ with $\bm\sigma_{\pm\theta}$, see Eq.~\ref{H_trilayer_staircase}, breaks the particle-hole symmetry $P$~\eqref{particle_hole}. The symmetry breaking is negligible in the small twist angle regime $\theta$ and has been ignored in the previous analysis. 
In the following we first discuss the stability of the zero modes to this perturbation. Then, we move to consider the effect of the perturbation away from the chiral limit.

\subsection{Robustness of the flat bands}
\label{subsec:robustness}

In the chiral limit the twist angle difference between top and bottom layer $\bm\sigma_{\pm\theta}$ enters in the Hamiltonian $ H_{\rm ABA}$~\eqref{H_ABA} as: 
\begin{equation}
\label{matrix_operators_compact_twist}
 \mathcal D(\br) =  -i\sqrt{2} M_{-\theta} \partial + \mathcal A(\br),  
\end{equation}
where $M_\theta=\diag\left(e^{i\theta},1,e^{-i\theta}\right)$. Differently from twisted bilayer graphene~\cite{Grisha_TBG} the layer dependent phases $M_\theta$ in Eq.~\eqref{matrix_operators_compact_twist} cannot be gauged away since $[M_{\theta/2} ,\mathcal A(\br)]\neq0$.
As mentioned before, the layer dependent phase breaks the particle-hole symmetry $P$~\cite{Yuncheng2023,devakul2023magicangle} while $C_{3z}$ and $C_{2y}T$ are still symmetries of $\mathcal H_{\rm ABA}$. In the chiral limit $w_{\rm AA}=0$ the Hamiltonian $\mathcal H_{\rm ABA}$ is characterized by three Dirac cones at $K$, $K'$ and $\Gamma$ that are protected by $C_{3z}$ and the chiral symmetry $\Lambda_z$. 
From Eq.~\eqref{matrix_operators_compact_twist} we readily realize that the orthogonality relation~\eqref{orthogonality_relation} transforms into the identity: 
\begin{equation}
\label{generalized_orthogonality_relation}
    v_\theta(\br) = \bar{\bm\chi}_{\bk_1}(\br)\cdot \left[M_\theta\bm\psi_{\bk_2}(\br)\right]=0
\end{equation}
for $\bk_1\neq\bk_2$ and arbitrary $\br$. We emphasize that this is no longer a vector product as it is not definite positive. Relations~\eqref{generate_chiral} and~\eqref{generate_antichiral} are also modified accordingly and become $\bm\psi_{-\bk_1-\bk_2}=[M_{\theta}\bar{\bm\chi}_{\bk_1}]\times [M_{\theta}\bar{\bm\chi}_{\bk_2}]$ and $\bm\chi_{-\bk_1-\bk_2}=[M_{-\theta}\bar{\bm\psi}_{\bk_1}]\times [M_{-\theta}\bar{\bm\psi}_{\bk_2}]$, while Eqs.~\eqref{wronskian_A} and~\eqref{wronskian_B} are still satisfied. 

At odd magic angles, the linear vanishing of $\bm\chi_\Gamma(\br)$ around $\br=0$ still holds, protected by the symmetries $C_{3z}$ and $C_{2y}T$. The positions of the magic angles are only slightly shifted by the particle-hole symmetry breaking term, $\alpha_1^* \approx 0.3772$ for instance. Moreover, since  
\begin{equation}
    \bar{\bm \chi}_\Gamma (0) = 0 =  [M_{\theta} {\bm\psi}_{K}]\times [M_{\theta} {\bm\psi}_{K'}],
\end{equation}
we still obtain that $ \bm\psi_K(0)=\bm\psi_{K'}(0)$ at the odd magic angle, yielding the Chern $2$ zero-mode chiral solution of Eq.~\eqref{flatAband} together with the anti-chiral Chern $1$ band. The structure of the flat bands are thus fully stable under particle-hole symmetry breaking as confirmed by our numerical calculations.

At even magic angles, as discussed in Appendix~\ref{app:breaking-P}, the $C_{3z}$ and $C_{2y}T$ symmetries alone are sufficient to preserve a double zero in one of the two wavefunctions $\bm\psi_\Gamma(\br)$ at $\br=0$, which automatically yields the two flat bands in the anti-chiral sector. The analytical expressions for these bands are explicitly given in Eq.~\eqref{analytical_solution_ABA_3}. It is worth noting that the choice of $\bm\psi_\Gamma(\br)$ vanishing at $\br=0$ is not continuously connected to the zero-mode solution $\bm\psi_\Gamma$ away from magic angles; instead, it requires the admixture with the first excited band, which occurs at even magic angles. In the anti-chiral sectors, the two flat bands are protected by the conservation of ${\rm Tr} \Lambda^z=0$ (see Eq.~\eqref{chiral_operator}) within the zero-energy manifold. This conservation condition dictates that the two chiral flat bands must be accompanied by two anti-chiral flat bands. Consequently, we find that the fourfold degenerate flat band structure is maintained at even magic angles, even in the presence of particle-hole symmetry breaking. Our numerical calculations confirm this protection. 
Finally, as already noted in Ref.~\cite{devakul2023magicangle}, the additional Dirac cone at $\Gamma$ is gapped by breaking $P$. The gap is however quite small, on the order of the energy scale $\sim \theta v_Fk_\theta$.

\subsection{Away from the chiral limit}
\label{subsec:chern num}

\begin{figure}[!ht]
    \centering
\includegraphics[width=\linewidth]{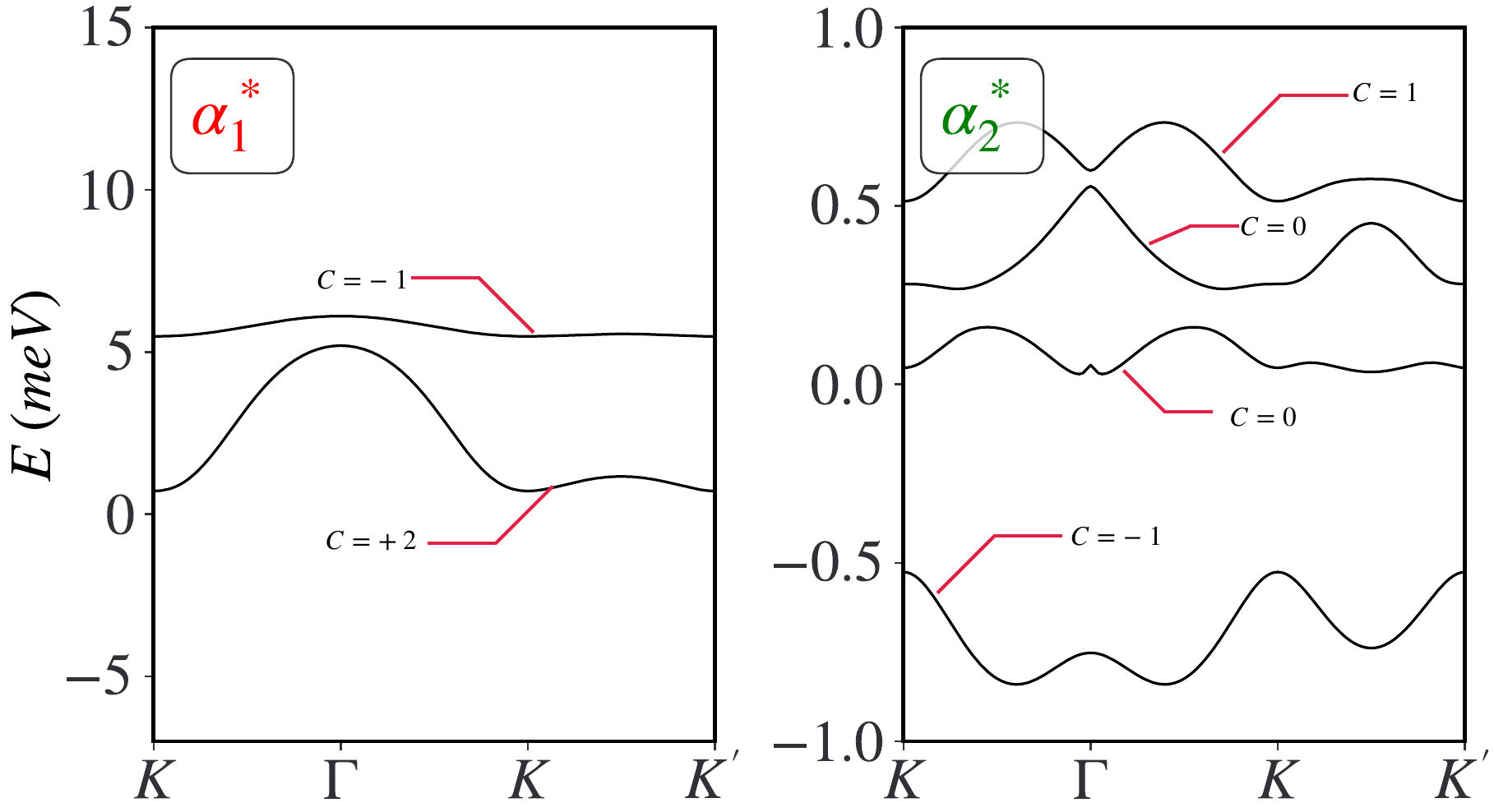}
    \caption{Separated middle bands of the first (left panel) and the second (right panel) magic angles, labeled with the Chern number associated to each band. The separated band are obtained by breaking simultaneously the particle-hole and the chiral symmetry. To amplify the effect of the particle-hole symmetry breaking the twist angle in $\bm\sigma_{\pm\theta}$ is multiplied by a factor 20.}
    \label{fig:chern num}
\end{figure}

The primary emphasis of this work has been on the chiral limit, which allowed us to provide analytical wavefunctions for the flat bands.
We now introduce a slight deviation from this limit by incorporating a finite corrugation factor $w_{AA}/w_{AB} = 0.05$, while still accounting for the particle-hole symmetry breaking term. In this case, all zero-energy modes are lifted and gaps open between bands as illustrated in Fig.~\ref{fig:chern num} in the vicinity of the first two magic angles. At twist angles different from magic values, the Dirac cones at $\Gamma$, $K$ and $K'$ become gapped when both the chiral and particle-hole symmetries are broken. At magic angles, whether even or odd, the flat bands acquires a finite bandwidth, on the order of a few meV for $w_{AA}/w_{AB} = 0.05$, and separate from each other.

As discussed in Sec.~\ref{subsec:odd_magic}, the odd magic angles in the chiral limit exhibit a flat band with Chern number $+2$ ($-1$) which is fully polarized on the A(B) sublattice. Remarkably, even when departing from the chiral limit, where the two sublattices become mixed, the resulting two bands still retain their Chern numbers of $+2$ and $-1$, as shown in Fig.\ref{fig:chern num} for the first magic angle. These Chern numbers are in fact robust and persist up to significant values of the corrugation as the gap to the remote bands is large.

In contrast to this, as discussed in Sec.~\ref{subsec:even_magic}, we identified a set of four flat bands with Chern numbers $+1, +1, -1, -1$, and complete sublattice polarization in the chiral limit for even magic angles. The Dirac cone crossing at $\Gamma$, as shown in the previous section, is gapped out by particle-hole symmetry breaking. When we move outside the chiral limit, the mixing of sublattices leads to the rearrangement of Chern numbers to $+1, 0, 0, -1$, as depicted in Fig.\ref{fig:chern num}, for the four split bands at a small corrugation of $w_{AA}/w_{AB} = 0.05$.
Increasing the corrugation further leads to topological transitions that involve the Dirac cone at $\Gamma$ and change the topological properties of the flat bands. Their study is beyond the scope of this work.

\section{Conclusions}
\label{sec:conclusions}

This work analyses the mathematical structure of flat bands in equal-twist helical trilayer graphene in the chiral limit and for ABA stacking. We determine analytical expressions for all zero-energy wavefunctions at magic angles together with the band Chern numbers.
We derive an orthogonality relation between chiral and anti-chiral zero modes, which constrains the number of generators in the zero-energy manifold and reveals a connection between the dimensionality of the vector space spanned by the zero modes and the total Chern number of the band.

In contrast to twisted bilayer graphene, helical twisted trilayer graphene exhibits an even/odd variation in the composition and features of the flat bands. At odd magic angles, we find a twofold degenerate zero-energy manifold at each $\bk$ point, comprising an anti-chiral flat band with Chern number $C_B=-1$ and a chiral flat band with Chern number $C_A=2$.
The latter, being generated by two linearly independent spinors, cannot be reduced to the lowest Landau level, leading to interesting implications on the properties of the correlated ground state~\cite{guerci2023chern}, which will be the topic of future studies.

Even magic angles, on the other hand, are distinguished by a four-fold degenerate manifold, where both chiral and anti-chiral flat bands have pairs of, or double, zeros, resulting in two zero-modes in each sector. The two sublattice-polarized bands in a given sector are all collinear to a single space-dependent spinor and carry the same Chern number, $C_A=+1$ for the chiral $A$-polarized bands, $C_B=-1$ for the anti-chiral $B$-polarized bands.
We also demonstrate that in addition to the two flat bands, there must be two additional degenerate zero modes within the zero-energy manifold, thereby explaining the presence of a Dirac cone crossing the flat bands at $\Gamma$.

We also investigate the stability of flat bands and zero modes at magic angles under a weak breaking of particle-hole symmetry. We find that all the features listed above remain valid, with the exception of the Dirac cone, which becomes slightly gapped by the perturbation. Only the joint breaking of particle-hole and chiral symmetries gaps out and splits all bands. Interestingly, at odd magic angle, the resulting isolated bands retain the Chern numbers $+2,-1$ analytically determined in the chiral limit. In the light of of Ref.~\cite{Stern_2023}, it is natural to ask which features protect the emergence and properties of the flat bands. Our analysis highlights the importance of the structure of the differential operator $\mathcal D =-i\sqrt{2} \partial + \mathcal A$, where $\mathcal A$ is a non-abelian traceless SU(3) gauge potential and there is clearly a natural generalization to the SU(N) case for multilayer stackings with $N>3$, where Chern bands with $C>2$ are expected.  In addition, the symmetries $C_{3 z}$ and $C_{2y}T$ appear to be  crucial for maintaining exactly flat bands tuned only by the twist angle, whereas particle-hole symmetry is not needed. We emphasize that, in contrast to twisted bilayer graphene, the $C_{2z}T$ symmetry is broken here and the model belongs to the Altland-Zirnbauer~\cite{altland1997} class AIII instead of CI, although both classes can have protected Dirac cone on a 2D surface.

Finally, the emergence of isolated bands with non-zero Chern values in a more realistic context, where both chiral and particle-hole symmetries are broken, opens up exciting possibilities for realizing an anomalous quantum Hall effect without the need for an almost aligned hBN substrate. Moreover, the potential for fractional Hall states consisting of bands with Chern numbers of $\pm 2$ 
represents a promising direction for future investigations.


\begin{acknowledgments}

We acknowledge discussions with Jie Wang, Jen Cano and Eslam Khalaf. C.M. and Y.M. acknowledge support by the French National Research Agency (project TWISTGRAPH, ANR-21-CE47-0018). D.G. acknowledges support from the Flatiron Institute, a division of the Simons Foundation.

\end{acknowledgments}

\appendix

\section{Local Hamiltonian}
\label{app:local_Hamiltonian}

We review here some fundamental properties of the local Hamiltonian describing equal twist helical trilayer graphene. A more general formulation, which includes non-equal twist configurations, is provided in Ref~\cite{Yuncheng2023,popov2023magic_butterfly}.
In the basis ${\bm\Psi}=(\psi_1,\chi_1,\psi_2,\chi_2,\psi_3,\chi_3)$ where $\psi,\,\chi$ correspond to the wave function amplitude on the A and B sublattices, respectively, the Hamiltonian near valley K reads~\cite{Yuncheng2023,guerci2023chern}
\begin{equation}
\label{H_trilayer_staircase}
        H_{\rm eTTG}(\br;{\bm \phi}) =\begin{pmatrix}
    v_F\hat{\bk}\cdot\bm \sigma_\theta & T(\br,{\bm \phi}) & 0\\
    h.c. & v_F\hat{\bk}\cdot\bm \sigma & T(\br,-{\bm \phi}) \\
    0 & h.c. & v_F\hat{\bk}\cdot\bm \sigma_{-\theta}
    \end{pmatrix},
\end{equation}
where $v_F\approx 10^6$m/s is the graphene velocity. The set of phases ${\bm \phi}=(\phi_1,\phi_2,\phi_3)$ parametrizes the position on the supermoiré lattice. With the choice of gauge $\phi_1=0$, 
\begin{equation}
   \bR = \frac{\phi_2}{\pi}{\mathbf a}^{\rm MM}_1+\frac{\phi_3}{\pi}{\mathbf a}^{\rm MM}_2,  
\end{equation}
with the supermoir\'e lattice vectors ${\mathbf a}^{\rm MM}_{1/2}=\frac{4\pi}{3\theta k_\theta} e^{\mp i\pi/3}$. ${\bm \phi}$ also controls the relative shift between the two moiré patterns and a change of gauge shifts all phases $\phi_j$ by the same amount. $\bR=0$ defines the AAA stacking, whereas ABA and BAB correspond to $\bR = ({\mathbf a}^{\rm MM}_2-{\mathbf a}^{\rm MM}_1)/3$ and $\bR = ({\mathbf a}^{\rm MM}_1-{\mathbf a}^{\rm MM}_2)/3$ parametrized by ${\bm \phi} = \pm (0,-\pi/3,+\pi/3)$.

${\bm \sigma}$ is the vector of Pauli matrices in the sublattice space, 
$\bm\sigma_\theta\equiv e^{i\theta\sigma^z/2}\bm\sigma e^{-i\theta\sigma^z/2}$ and $\hat\bk=-i\nabla_{\br}$. 
The tunneling between different layers is described by the moir\'e potential: 
\begin{equation}
    T(\br,{\bm \phi})=\sum^{3}_{j=1} T_j e^{-i\br\cdot\bq_j}e^{-i\phi_j},
\end{equation}
where $T_{j+1} = w_{\rm AA} \sigma^0 + w_{\rm AB}[\sigma^x\cos2 \pi j/3  + \sigma^y\sin 2 \pi j/3]$, $w_{\rm AB}=110$meV and $w_{\rm AA}=r w_{\rm AB}$ with $r$ dimensionless parameter quantifying atomic corrugation, using complex notation $\bq_{j+1}= ik_\theta e^{2i\pi j/3}$~\cite{Christophe_2019} with $k_\theta=\theta K_D$, $K_D=4\pi/3a_{\rm G}$ and $a_{\rm G}\approx2.46$\AA. The moir\'e lattice is characterized by the reciprocal lattice vectors $\bb_{1/2}=\bq_1-\bq_{2/3}$ and primitive vectors $\ba_{1/2}$. 
Ignoring the twist angle dependency in the Pauli matrices $\bm\sigma_{\pm\theta}$ the Hamiltonian~\eqref{H_trilayer_staircase} becomes invariant under the particle-hole symmetry: 
\begin{equation}
    P H_{\rm eTTG}(\br;{\bm \phi})  P^{-1}= - H_{\rm eTTG}(-\br;{\bm \phi}) , 
\end{equation}
with $P$ given in Eq.~\eqref{particle_hole}. Under moir\'e lattice translations we have: 
\begin{equation}
\label{H_boundary_conditions}
    H_{\rm eTTG}(\br+\ba_{1/2};\bm\phi) = U_\varphi H_{\rm eTTG}(\br;\bm\phi) U^\dagger_\varphi,
\end{equation}
where the matrix $U_{\varphi}=\diag[\omega^*,1,\omega]\otimes\sigma^0$ corresponds to a layer dependent phase factor $\omega=\exp (2\pi i/3)$. 

$H_{\rm eTTG}$ also exhibits a supermoiré periodicity with $\bR$ (or ${\bm \phi}$). One can show~\cite{guerci2023chern} the following identity
\begin{equation}
    H_{\rm eTTG}\left(\br+\frac{\ba_{l}}{2};{\bm \phi}+\Delta{\bm\phi}_{l}\right)=\tilde UH_{\rm eTTG}(\br;{\bm \phi})\tilde U^\dagger, 
\end{equation}
where $\tilde U=\diag(1,1,\omega)$. The phase shifts $\Delta {\bm\phi}_1 = (0,\pi,0)$, $\Delta {\bm\phi}_2 = (0,0,\pi)$ correspond respectively to $\bR \to \bR + {\mathbf a}^{\rm MM}_1$ and 
$\bR \to \bR + {\mathbf a}^{\rm MM}_2$.
As a result, the AAA stacking points are periodically replicated, forming a triangular lattice generated by ${\mathbf a}^{\rm MM}_{1/2}$ and characterized by ABA and BAB domains~\cite{guerci2023chern} (see also~\cite{devakul2023magicangle,nakatsuji2023multiscale}).

As noted above, we focus on the ABA stacking configuration by setting ${\bm\phi} = (0,2 \pi/3,-2 \pi/3)$ and the chiral limit $w_{AA}=0$ (suppressed tunneling between A and A orbitals). The resulting Hamiltonian is given by Eq.~\eqref{H_ABA} in the sublattice-Chern basis.

\section{Particle-hole symmetry $P$ protected Dirac cones}
\label{subsec:ph_protection}

We discuss the symmetries protecting the Dirac cones at the high-symmetry points $K$, $K'$ and $\Gamma$ in the ABA stacking configuration. 

The irreducible representations of the space group composed by $C_{3z}$ and $C_{2y}T$  at the high-symmetry momenta $\Gamma$ and $K$ ($K'$) verify the $C_{3}$ point group character table and are all one-dimensional. 
This can be understood more directly by considering an eigenstate of $C_{3z}$,  $C_{3z} | \omega \rangle = \omega | \omega \rangle$. From the relation
\begin{equation}
    (C_{2 y} T) C_{3 z} (C_{2 y} T)^{-1} = C_{3 z}^{-1}, 
\end{equation}
we obtain $C_{3 z} C_{2 y} T |\omega \rangle = \omega \, C_{2 y} T | \omega \rangle$. $C_{2 y} T$ thus does not circulate between the eigenstates of $C_{3z}$ and cannot protect a twofold degeneracy. The three zero-energy Dirac cones at $K$, $K'$ and $\Gamma$ arising in the band spectrum of ABA trilayer graphene~\cite{Yuncheng2023,guerci2023chern} are therefore stable in the presence of the particle-hole symmetry $P$~\eqref{particle_hole}. In the chiral limit $w_{\rm AA}=0$, the Dirac cones are protected by $\Lambda^z$ and persists even in the absence of $P$.

Since $P$  and $C_{3 z}$ commute, if the spectrum at $\Gamma$ hosts two states with $C_{3z}$ eigenvalues $\omega$, $\omega^*$ near charge neutrality (and no other states), particle-hole symmetry $P$ automatically pins these two states at zero energy. This is proven by contradiction: if we assume that the two states sit at opposite non-vanishing energies, then $P$ permutes them. This is however impossible since $P$ cannot change the $C_{3z}$ eigenvalue which completes the proof. In fact, $P$ restricted to these two states must be the identity as it commutes with $C_{3 z}$. It further shows that breaking $C_{3 z}$ does not lift the Dirac crossings as the trivial (identity) representation of $P = I_{2\times2}$ cannot deform continuously to the traceless $\sigma^x$ matrix permuting states with opposite non-zero energies. $K$ and $K'$ are however not stable under $P$ - which permutes $K$ and $K'$ - and the stability of their Dirac cones must come from a different operator. $C_{2y}T$ sends $\bk\to-C_{2y}\bk$ and therefore also permutes $K$ and $K'$. Combining $P$ and $C_{2y}T$ we find:
\begin{equation}
    PC_{2y}T\mathcal H_{\rm ABA}(\br) \left(PC_{2y}T\right)^{-1} = -\mathcal H_{\rm ABA}(-C_{2y}\br),
\end{equation}
with 
\begin{equation}
    P'=PC_{2y}T =\begin{pmatrix}
        1 & 0 & 0 \\
        0 & -1 & 0 \\
        0 & 0 & 1
    \end{pmatrix}\otimes I_{2\times 2}\mathcal K,
\end{equation}
which leaves $K$ and $K'$ invariant and acts as a (anti-unitary) particle-hole operator. $P'C_{3z}(P')^{-1}=C^{-1}_{3z}$ being anti-unitary $P'$ cannot permute the eigenvalues $(\omega,\omega^*)$ which implies that an isolated pair of states at $K$ and $K'$ is degenerate and pinned at zero energy, in the two-dimensional subspace $P'=I_{2\times 2}\mathcal K$.

\section{Symmetries of the zero-mode wavefunction}
\label{app:symmetries_zeromode}

The symmetries of the Hamiltonian $\mathcal H_{\rm ABA}$ discussed in Section~\eqref{subsec:symmetries} constraints the properties of the zero-mode wavefunction and gives insight on the nature of the different magic angles. Considering the symmetry $g$ the A sublattice polarized solution $\bm\psi$ we have:
\begin{equation}
\label{symmetry_wfc}
   \bm \psi_{\bk}(\br) = e^{i\Xi_{g}}M_g \bm\psi_{g\bk}(g\br),
\end{equation} 
where the action of the symmetry $g$ on layer, momentum and space degrees of freedom is given in table~\ref{tab:Symmetries_ABA}. 
\begin{table}[]
\centering
\begin{tabular}{|c||c|c|c|}
\hline
Symmetry &  $\bk$ & $\br$ & $M_g$   \\
\hline\hline
$C_{3z}$ & $C_{3z}\bk$ & $ C_{3z}\br$ & 
$\diag(\omega,1,\omega)$\\
\hline
$ P$ & $-\bk$ & $-\br$ & $ \begin{pmatrix}
        0 & 0 & -1 \\
        0 & 1 & 0 \\
        -1 & 0 & 0
    \end{pmatrix}$ \\
\hline
\hline
$C_{2y}T$ & $-C_{2y}\bk$ & $C_{2y}\br$ & $ \begin{pmatrix}
        0 & 0 & 1 \\
        0 & 1 & 0 \\
        1 & 0 & 0
    \end{pmatrix}\mathcal K$ \\
\hline
$PC_{2y}T$ & $C_{2y}\bk$ & $-C_{2y}\br$ & 
    $\diag(-1,1,-1)\mathcal K$\\
\hline
\end{tabular}
\hspace{10.cm}
\caption{Action of the symmetries of $\mathcal H_{\rm ABA}$ on the zero-mode $\bm\psi$. Rows refer to the different symmetries, from top to bottom $C_{3z}$, $P$, $C_{2y}T$ and $PC_{2y}T$ the last two involving the complex conjugation $\mathcal K$. Columns show the action of the symmetry on the momentum $\bk$, space $\br$ coordinates and $M_g$ is the representation of the symmetry acting on the three dimensional layer degree of freedom.
} 
\label{tab:Symmetries_ABA}
\end{table}
Similar expressions are also obtained for the B sublattice zero mode $\bm\chi_{\bk}$. The phase $\Xi_{g}$ is fixed by taking the decoupled limit $\alpha=0$ and depends on $\bk$.

We will now employ the symmetry relations in Eq.~\eqref{symmetry_wfc} of the model to constrain the zero mode wavefunctions in the ${\cal C}_A$ and ${\cal C}_B$ sectors around high-symmetry points $\br$ in the moir\'e unit cell.  

\subsection{Odd magic angles}
\label{app:odd_magic}

Thanks to Eq.~\eqref{magic_relation} odd magic angles are determined by looking at the behavior of $\bm\chi_\Gamma(\br)$ around $\br\approx0$ where the kernel $\mathcal D(\br)$ takes the form: 
\begin{equation}
\label{expansion_around_AA}
    \mathcal D(\br)\simeq \begin{pmatrix}
    -i\sqrt{2}\partial & -3\alpha z/\sqrt{2} & 0 \\
    3\alpha & -i\sqrt{2}\partial & 3\alpha  \\
    0 & 3\alpha z/\sqrt{2} & -i\sqrt{2}\partial 
\end{pmatrix} .  
\end{equation}
The solution of the zero-mode equation is obtained performing a Taylor expansion in $z$ and $\bar z$. Employing Eq.~\eqref{symmetry_wfc} we readily realize that $\chi_{\Gamma 2}(C_{3z}\br)=\chi_{\Gamma2}(\br)$ and $\chi_{\Gamma 1/3}(C_{3z}\br)=\omega^*\chi_{\Gamma 1/3}(\br)$. In addition we also have $\chi_{\Gamma 2}(C_{2y}\br)=\chi^*_{\Gamma 2}(\br)$ and $\chi_{\Gamma 1/3}(C_{2y}\br)=\chi^*_{\Gamma 3/1}(\br)$ implying:
\begin{equation}
\label{second_order_expansion}
 \bm  \chi_\Gamma(\br)\simeq\chi_{ 2} \begin{pmatrix}
     3i\alpha z^2/4  \\
     1  \\
     -3i\alpha z^2/4 
 \end{pmatrix} +  \begin{pmatrix}
      \chi'_{1} \bar z  \\
      3\sqrt{2}\alpha \Im\chi'_{1} z\bar z \\
      {\chi'}^*_{1} \bar z 
 \end{pmatrix} ,
\end{equation}
where $\chi_2\equiv\chi_{\Gamma 2}(0)\in\mathbb R$ while $\chi'_1\equiv\bar\partial\chi_{\Gamma 1}|_0\in\mathbb C$. Thus, $C_{3z}$ and $C_{2y}T$ reduces  the perfect flatness of the entire band to the vanishing of a single real number $\chi_2$~\cite{Stern_2023}.
Notice that if we further impose the particle-hole symmetry $P$ we have $\Re\chi'_1=0$. We readily realize that the expression~\eqref{second_order_expansion} solves $\mathcal D\bm\chi_\Gamma=0$ up to small terms of the order $z^2\bar z$. At the magic angle we have $\chi_{2}=0$ and $\bm\chi_{\Gamma}$ has a simple zero for $\br\to0$: 
\begin{equation}
\label{Taylor_magic_AA}
\bm\chi_{\Gamma}(\br\to0)\sim \bar z\left(\chi'_1,0,{\chi'}^*_1\right)^T.
\end{equation} 
We emphasize that the simple zero at the odd magic angles where $\chi_2=0$ persists also in the absence of the particle-hole symmetry $P$. The vanishing of $\bm\chi_\Gamma(0)$ implies $\bm\psi_{K}(0)\times\bm\psi_{K'}(0)=0$ resulting in the fact that $\bm\psi_K(0)$ and $\bm\psi_{K'}(0)$ are colinear. Eq.~\eqref{symmetry_wfc} for $C_{3z}$ gives $\psi_{K/K'2}(C_{3z}\br)=\omega \psi_{K/K'2}(\br)$ and $\psi_{K/K'1/3}(C_{3z}\br)= \psi_{K/K'1/3}(\br)$. Furthermore, $K$ and $K'$ zero modes are related by $\psi_{K'1/3}(\br)=\psi^*_{K3/1}(C_{2y}\br)$ and $\psi_{K'2}(\br)=\psi^*_{K2}(C_{2y}\br)$. These symmetries reduces $\bm\psi_{K}(0)\times\bm\psi_{K'}(0)=0$ to $\bm\psi_{K}(0)=\bm\psi_{K'}(0)$, where the identity holds up to a phase, which gives rise to the the Chern 2 zero mode wavefunction in Eq.~\eqref{flatAband}.

\subsection{Even magic angles}
\label{app:even_magic}

Even magic angles realizes the a 1+1 decomposition corresponding to one-dimensional flat bands in both chiral and anti-chiral sectors. The four-fold degeneracy of the flat band manifold, see Fig.~\ref{fig:flatbands_1st_2nd}c, originates from a double zero in the spinors $\bm\psi_{K'}(\br)$ and $\bm\chi_{K}(\br)$ at $\br_0$. The origin of this double zero can be explained by symmetry reasoning. To start with we expand $\mathcal D^\dagger(\br)$ around the AB stacking point $\br_0$ finding:
 \begin{equation}
    \mathcal D^\dagger(\br+\br_0)\simeq \begin{pmatrix}
    -i\sqrt{2}\bar\partial & -3\alpha z/\sqrt{2} & 0 \\
    3\alpha z/\sqrt{2} & -i\sqrt{2}\bar\partial & 3\alpha  \\
    0 & -3\alpha \bar z/\sqrt{2} & -i\sqrt{2}\bar\partial 
\end{pmatrix}.
\end{equation}
Focusing on the chiral sector and fixing the center of the $C_{3z}$ rotation around $\br_0$ we find Eq.~\eqref{C3z_r0} which implies: 
\begin{equation}
    \bm\psi_{K'}(\br+\br_0)\simeq\psi^{(0)}_3\begin{pmatrix}
    0  \\
     -i3\alpha\bar z/\sqrt{2}  \\
     1 
 \end{pmatrix} + \frac{z^2}{2} \begin{pmatrix}
      \psi^{(0)''}_1  \\
      \psi^{(0)''}_2 \\
      0
 \end{pmatrix}  ,
\end{equation}
where $\psi^{(0)''}_{1/2}\equiv\partial^2\psi_{K'1/2}|_{\br_0}$ and we have included terms up to second order in $\br$. Imposing, the $P C_{2y}T$ symmetry~\eqref{PC2yT_r0} implies $\Re\psi^{(0)}_3=0$, $\Re\psi^{(0)''}_1=0$ and $\Im\psi^{(0)''}_2=0$. Thus, at the magic angle where $\psi^{(0)}_3=0$ we have 
\begin{equation}
    \bm\psi_{K'}(\br\to0+\br_0)\sim z^2\left(\psi^{(0)''}_1 ,\psi^{(0)''}_2 ,0\right)^T/2,
\end{equation}
enabling to attach two lowest Landau levels with simple pole at $\br_0$ to the zero mode spinor $\bm\psi_{K'}(\br)$~\eqref{analytical_solution_ABA_2}.   
Following a similar line or reasoning one could show that the anti-chiral zero mode $\bm\chi_{K}$ shows a double zero at $\br_0$.

\subsection{Alternative derivation for even magic angles}
\label{app:breaking-P}

We provide here an alternative argument for the protection of a twofold degenerate flat band at even magic angles. The construction in Sec.~\ref{subsec:even_magic} relies on the quadratic vanishing of $\bm\psi_{K'}(\br)$ at $\br = \br_0$ derived in Sec.~\ref{app:even_magic} which uses the $P C_{2y}T$ symmetry, and therefore $P$. However, the zero-mode flat bands also host a specific wavefunction $\bm\psi^{(1)}_{\Gamma}(\br) =\eta^{(0)}_{K'}(z)\eta^{(0)}_{K'}(z)\bm\psi_{K'}(\br)$ which exhibits a double zero at $\br = 0$. As we show below, this double zero is protected solely by the $C_{3z}$ and $C_{2y}T$ symmetries. To linear order in $z$, the zero mode equation takes the form:
\begin{equation}
\begin{split}
    &-i\sqrt{2}\bar\partial \psi_{\Gamma 2}-3\alpha\bar z(\psi_{\Gamma 1}-\psi_{\Gamma 3})/\sqrt{2},\\
    &-i\sqrt{2}\bar\partial(\psi_{\Gamma 1}-\psi_{\Gamma 3})=0,\\
    &-i\sqrt{2}\bar\partial(\psi_{\Gamma 1}+\psi_{\Gamma 3})+6\alpha\psi_{\Gamma 2}=0.
\end{split}
\end{equation}
Solving the system of differential equations and imposing $C_{3z}$, Eq.~\eqref{symmetry_wfc}, we find that the zero mode behaves as: 
\begin{equation}
    \bm\psi_\Gamma(\br)\simeq \psi_2\begin{pmatrix}
    0  \\
    1  \\
     0 
 \end{pmatrix} -\frac{3\alpha i\bar z}{\sqrt{2}}\psi_2 \begin{pmatrix}
    1  \\
    0  \\
    1 
 \end{pmatrix}+z^2\begin{pmatrix}
    \psi''_1/2  \\
    0  \\
    \psi''_3/2 
 \end{pmatrix}
\end{equation}
where $\psi_2\equiv\psi_{\Gamma 2}(0)$ and $\psi''_{ 1/3}=\partial^2\psi_{\Gamma 1/3}|_0$. $C_{2y}T$ gives the additional condition $\psi_{\Gamma 1/3}(C_{2y}\br) = \psi^*_{\Gamma 3/1}(\br)$ and $ \psi_{\Gamma 2}(C_{2y}\br) = \psi^*_{\Gamma 2}(\br)$ implying $\psi_{2}\in\mathbb R$ and $\psi''_1={\psi''}^*_3$. In summary, a double zero is expected,
\begin{equation}
\label{second_quadratic}
    \bm\psi_{\Gamma}(\br\to0)\sim z^2\left(\psi^{''}_1 ,0,{\psi''}^*_1 \right)^T/2
\end{equation}
as soon as the real coefficient $\psi_2$ vanishes. It corresponds to the specific solution $\bm\psi^{(1)}_{\Gamma}(\br)$ introduced above. Breaking $P$ but keeping $C_{3z}$ and $C_{2y}T$ intact, one only moves the magic angle for which $\psi_2=0$ since $\psi_2$ remains a real number. The twofold degenerate flat band in the chiral sector is a direct consequence of Eq.~\eqref{second_quadratic}, with the analytical structure
\begin{equation}
\label{analytical_solution_ABA_3}
    \bm\psi_{\bk}(\br) = \eta^{(0)}_{\bk'}(z)\eta^{(0)}_{\bk-\bk'}(z) \bm\psi_{\Gamma} (\br),
\end{equation}
alternative to Eq.~\eqref{analytical_solution_ABA_2}. 

\bibliography{sample}

\end{document}